\documentclass[fleqn,usenatbib]{mnras}

\usepackage[T1]{fontenc}
\usepackage{ae,aecompl}
\usepackage{bm}
\usepackage{graphicx}    

\usepackage{amsmath}	% Advanced maths commands
\usepackage{amssymb}	% Extra maths symbols

\usepackage{txfonts}

\title{Orbital dynamics of circumbinary planets }

\author[Chen et al.]{Cheng Chen$^1$\thanks{Email: chenc21@unlv.nevada.edu}, Alessia Franchini$^1$,  Stephen H. Lubow$^2$ and Rebecca G. Martin$^1$
\\ $^1$Department of Physics and Astronomy,  University of Nevada, Las Vegas, 4505 South Maryland Parkway, Las Vegas, NV 89154, USA 
\\ $^{2}$Space Telescope Science Institute, 3700 San Martin Drive, Baltimore, MD 21218, USA\\}

\date{Accepted XXX. Received YYY; in original form ZZZ}

\pubyear{2019}
\begin{document}
\label{firstpage}
\pagerange{\pageref{firstpage}--\pageref{lastpage}}
\maketitle

% Abstract of the paper
\begin{abstract}
We investigate the dynamics of a nonzero mass, circular orbit planet around an eccentric orbit binary for various values of the binary eccentricity, binary mass fraction, planet mass, and planet semi--major axis by means of numerical simulations. Previous studies investigated the secular dynamics mainly by approximate analytic methods. In the stationary inclination state,  the planet and binary precess together with no change in relative tilt. For both prograde and retrograde planetary orbits, we explore the conditions for planetary orbital libration versus circulation and the conditions for stationary inclination. As was predicted by analytic models, for sufficiently high initial inclination, a prograde planet's orbit librates about the stationary tilted state. For a fixed binary eccentricity, the stationary angle is a monotonically decreasing function of the ratio of the planet--to--binary angular momentum $j$. The larger $j$, the stronger the evolutionary changes in the binary eccentricity and inclination. We also calculate the critical tilt angle that separates the circulating from the librating orbits for both prograde and retrograde planet orbits. The properties of the librating orbits and stationary angles are quite different for prograde versus retrograde orbits. The results of the numerical simulations are in very good quantitative agreement with the analytic models. Our results have implications for circumbinary planet formation and evolution.
\end{abstract}


\section{Introduction}

Binary stars form in turbulent molecular clouds \citep{McKee2007} where the accretion process during star formation may be chaotic \citep{Bate2003,Bate2018}. Therefore circumbinary discs likely form misaligned with respect to the binary orbital plane.  Observationally, misaligned circumbinary discs appear common  \citep[e.g.,][]{Brinch2016,Chiang2004,Winn2004,Kennedy2012,Kennedy2019}. 
Giant planets form while the gas disc is still present \citep[e.g.,][]{Lagrange2010} and thus it is likely that planets may form on misaligned orbits.  A giant planet in a misaligned disc in a binary system may not remain coplanar to the disc  \citep{Picogna2015,Lubow2016,Martin2016,Franchini2019c}. 

In this paper we concentrate on the properties
 of circumbinary planets involving eccentric orbit binaries.
The Kepler Mission has so far detected  10 circumbinary planets. 
Among these, at least two of them have been detected around eccentric binaries. Kepler-34b with a mass of $0.22\,M_{\rm J}$ orbits an eclipsing binary star system (Kepler-34) that has an orbital eccentricity of 0.52 
\citep{Welsh2012,Kley2015}. The moderately eccentric ($e=0.26$) binary KIC 5095269 hosts a circumbinary planet KIC 5095269b \citep{Getley2017}.

All the circumbinary planets detected this far are  nearly coplanar with the binary orbital plane. However, this is likely a selection effect due to the small orbital period of the Kepler binaries \citep{Czekala2019}. Longer orbital period binaries are then expected to host planets with  a wide range of  inclinations. Planets with large misalignments are much more difficult to detect than those that orbit in the binary orbital plane because their transits are much more rare.
However,  other detection methods may be possible. 
For example, polar planets (those in an orbit close to perpendicular to the binary orbital plane) may be distinguished from coplanar planets through eclipse timing variations \citep{Zhang2019}.

We consider circular circumbinary planet orbits throughout this paper.
For a misaligned test (massless) particle circumbinary orbit about a circular orbit binary, its nodal precession occurs around the binary angular momentum vector, but may be either prograde or retrograde depending upon the initial particle inclination.  The nodal precession about a circular orbit binary is always circulating. That is, the longitude of the ascending node fully circulates over $360^\circ$, since the nodal precession rate does not change sign. 
However for a binary with nonzero eccentricity, a circumbinary test  particle orbit with a sufficiently large inclination may undergo libration.  That is, the longitude
of the ascending node covers a limited range of angles less than $360^\circ$ and the nodal precession rate changes sign. 
In the test particle case, the angular momentum vector of the test particle librates about the binary eccentricity vector (or binary semi-major axis) and undergoes
tilt oscillations
\citep{Verrier2009,Farago2010,Doolin2011,Naoz2017,deelia2019}. The  minimum inclination required for libration decreases with increasing binary eccentricity. This means that a test particle orbit with even a small inclination can librate around a highly eccentric binary.

Misaligned low mass/angular momentum discs  in and around around binaries can undergo similar precession as a test particle
\citep{Larwoodetal1996}. Consequently, following the behaviour of test particles, a sufficiently misaligned low mass disc around an eccentric binary can precess around the eccentricity vector, rather than the binary angular momentum and undergo tilt oscillations \citep{Aly2015, MartinandLubow2017, MartinandLubow2018b, Lubow2018, Zanazzi2018,Franchini2019b}. Dissipation within the disc leads to the eventual alignment, either coplanar or polar aligned with respect to the binary orbital plane.  In the polar aligned state, the low angular momentum circumbinary disc lies perpendicular to the binary orbital plane and its angular momentum vector
is along the binary eccentricity vector. The polar disc as well as the binary do not undergo nodal precession. In the eccentric orbit binary, a disc that is evolving towards coplanar alignment undergoes tilt oscillations as it does so \citep{Smallwood2019}.

The mass of the disc has a significant effect on the polar alignment. A disc with mass is expected to
evolve to a generalised polar state  in which the inclination
of the disc relative to the binary is stationary in a frame that precesses with the binary.
 A simplified model for the disc with mass involves the orbit of a particle with mass. The
particle orbit then corresponds to a ring or narrow disc.
In this stationary
state, the disc inclination is less than $90^\circ$ for a disc in a prograde orbit \citep{Zanazzi2018,MartinandLubow2019}.

 In this work, we extend the three--body numerical simulations of a circular orbit, test (massless) circumbinary particle performed by \cite{Doolin2011} by allowing the particle (planet) to have nonzero mass. There has been some exploration into the stability of low mass polar particles \citep{Cuello2019}.
In this case, the binary feels the gravitational force of the planet which causes the binary orbit to be modified. The eccentricity vector of the binary precesses as a result of this interaction.   The  binary
orbit tilt and the magnitude of its eccentricity  oscillate in this case. 
 We consider the third body (planet) to lie on a circular circumbinary orbit. The dynamics of a circular orbit planet are similar to those of a narrow ring that in turn may be indicative of a more extended disc with mass.
Therefore, the results can also have implications for the evolution of a circumbinary disc with nonzero mass.
Our simulations are scale free and can therefore be applied to all scales. %However, in this work, we focus on the application of our model to planetary systems. Note: Galactic systems are not Keplerian.

In Section~\ref{results} we describe the initial conditions for our three--body simulations and explore the properties of circumbinary orbits with a planet, represented as a particle with nonzero mass. In Section~\ref{critical} we determine the critical angles between the librating and circulating orbits and compare them to the analytic solutions found by \cite{MartinandLubow2019} that is based on earlier work by \cite{Farago2010}. We also determine stationary inclinations, the inclination at which the binary and the third body orbit precess at the same rate with no change in relative tilt.  In Section~\ref{conclusion} we present our discussion and conclusions.

\section{Three--body simulations} 
\label{results}

In this section we describe and present results of our three--body simulations.  We explore the effects of the binary  eccentricity and the binary  mass ratio on the different orbit types, circulating and librating, for planets with varying angular momentum.

\begin{table}
\caption{Parameters of the simulations. The first column contains the name of the model, the second and third columns indicate the binary mass fraction and initial eccentricity. The fourth and fifth columns represent the mass of the planet in units of $m_{\rm b}$ and the distance of the planet with respect to the centre of the mass in units of $a_{\rm b}$ respectively. }
\centering
%\label{tab:example_table}
\begin{tabular}{ccccc} 
\hline
\textbf{Model} & $f_{\rm b}$ & $e_{\rm b}$ & $m_{\rm p}/m_{\rm b}$ & $r/a_{\rm b}$\\
\hline
\hline
A1 &  0.5  & 0.2 & 0.001 & 5 \\
A2 &  0.5  & 0.5 & 0.001 & 5  \\
A3 &  0.5  & 0.8 & 0.001 & 5 \\
B1 &  0.1  & 0.2 & 0.001 & 5 \\
B2 &  0.1  & 0.5 & 0.001 & 5  \\
B3 &  0.1  & 0.8 & 0.001 & 5 \\
\hline
C1 &  0.5  & 0.2 & 0.01 & 5 \\
C2 &  0.5  & 0.5 & 0.01 & 5  \\
C3 &  0.5  & 0.8 & 0.01 & 5 \\
D1 &  0.1  & 0.2 & 0.01 & 5 \\
D2 &  0.1  & 0.5 & 0.01 & 5  \\
D3 &  0.1  & 0.8 & 0.01 & 5 \\
\hline
E1 &  0.5  & 0.2 & 0.01 & 20 \\
E2 &  0.5  & 0.5 & 0.01 & 20  \\
E3 &  0.5  & 0.8 & 0.01 & 20 \\
F1 &  0.1  & 0.2 & 0.01 & 20 \\
F2 &  0.1  & 0.5 & 0.01 & 20  \\
F3 &  0.1  & 0.8 & 0.01 & 20 \\
\hline
G1 &  0.5  & 0.2 & 0.056 & 20  \\
G2 &  0.5  & 0.5 & 0.050 & 20 \\
G3 &  0.5  & 0.8 & 0.034 & 20 \\
H1 &  0.5  & 0.2 & 0.116 & 20 \\
H2 &  0.5  & 0.5 & 0.102 & 20 \\
H3 &  0.5  & 0.8 & 0.070 & 20 \\
\hline
\end{tabular}
\label{table1}
\end{table}

\subsection{Three--body simulation set--up}

To study the evolution of a third body orbiting around an eccentric binary star system, we use the $N$-body simulation package, REBOUND. 
We use a WHfast integrator which is a second order symplectic Wisdom Holman integrator with 11th order symplectic correctors \citep{Rein2015b}. 
We solve the gravitational equations for the three bodies
in the frame of the centre of mass of the three--body system.
The central binary has components of mass $m_1$ and $m_2$ with total mass $m_{\rm b}=m_1+m_2$ and the mass fraction of the binary is $f_{\rm b}=m_2/m_{\rm b}$. The orbit has semi--major axis $a_{\rm b}$, the magnitude of the eccentricity of the binary is $e_{\rm b}$ and the orbital period of the binary is $T_{\rm b}$. The  Keplerian orbit of the planet with mass $m_{\rm p}$ around the centre of mass of the binary is defined initially by six orbital elements, its semi-major axis $a$, inclination $i$, eccentricity $e$, longitude of the ascending node $\phi$, argument of periapsis $\omega$, and true anomaly $\nu$.  Since the planet orbit is initially circular, we set as initial conditions $e=0$ and $\omega=0$. We take $\nu$=0 and $\phi=90^{\circ}$ initially in our suites of simulations.
We note that the planet orbit remains nearly circular in our simulations, as is expected analytically  since
the particle eccentricity is a constant of motion in the secular quadrupole  approximation for the binary
\citep{Farago2010}. We vary the planet mass, initial inclination and semi-major axis of its orbit.  Note that  the
binary orbit is not fixed since the binary feels the gravity of the
massive third body.

In order to plot the results, we work in a frame defined by the instantaneous values of the
eccentricity and angular momentum vectors of the binary, $\bm{e}_{\rm b}$ and $\bm{l}_{\rm b}$ respectively.  
The frame defined by the binary has the three axes
$\bm{e}_{\rm b}$, $\bm{l}_{\rm b}\times \bm{e}_{\rm b}$, and $\bm{l}_{\rm b}$. Denoting the
planet angular momentum as $\bm{l}_{\rm p}$, we determine the inclination of its orbital plane relative to the binary through
\begin{equation}
i =\cos^{-1}(\bm{\hat l}_{\rm b}\cdot \bm{\hat l}_{\rm p}),
\end{equation}
where $\bm{\hat l}_{\rm b}$ is a unit vector in the direction of the
angular momentum of the binary and $\bm{\hat l}_{\rm p}$ is a unit
vector in the direction of the angular momentum of the
particle.
The inclination of the binary relative to the total angular momentum $\bm{l}$ is
\begin{equation}
i_{\rm b} =\cos^{-1}(\bm{\hat l}\cdot \bm{\hat l}_{\rm b}),
\end{equation}
where $\bm{\hat l}$ is a unit
vector in the direction of the total angular momentum ($\bm{l}=\bm{l}_{\rm b}+\bm{l}_{\rm p}$). Similarly, we determine the phase angle of the particle in the
same frame of reference through
\begin{equation}
\phi=\tan^{-1}\left(\frac{\bm{l}_{\rm p}\cdot (\bm{l}_{\rm b}\times \bm{e}_{\rm b})}{\bm{l}_{\rm p}\cdot \bm{e}_{\rm b}}  \right).
\end{equation}

In the next three subsections, we vary masses and orbital properties of the binary stars and planets and describe the binary and planet orbital evolution.
We first ran some test particle simulations with the planet mass set to zero in order to verify that our results are in agreement with the results presented in Figure 2 in \citet{Doolin2011}. 
We then performed a set of simulations for various values of the mass of the planet from $m_{\rm p}=0.001\,m_{\rm b}$ to $m_{\rm p}=0.01\,m_{\rm b}$ and its orbital radius from $r=5\,a_{\rm b}$ to $r=20\,a_{\rm b}$.  We also explored the effect of different initial binary eccentricities and mass ratios. We also consider some simulations with higher angular momentum of the third body. In Table~\ref{table1}, we list the parameters for each model. 
%{\bf We limit our attention to circular orbit planets. Such orbits are applicable also to models of a narrow circular ring that may serve as a model for a disc.} 

\subsection{Low mass planet at small orbital radius}

\begin{figure*}
  \centering
    \includegraphics[width=17.7cm]{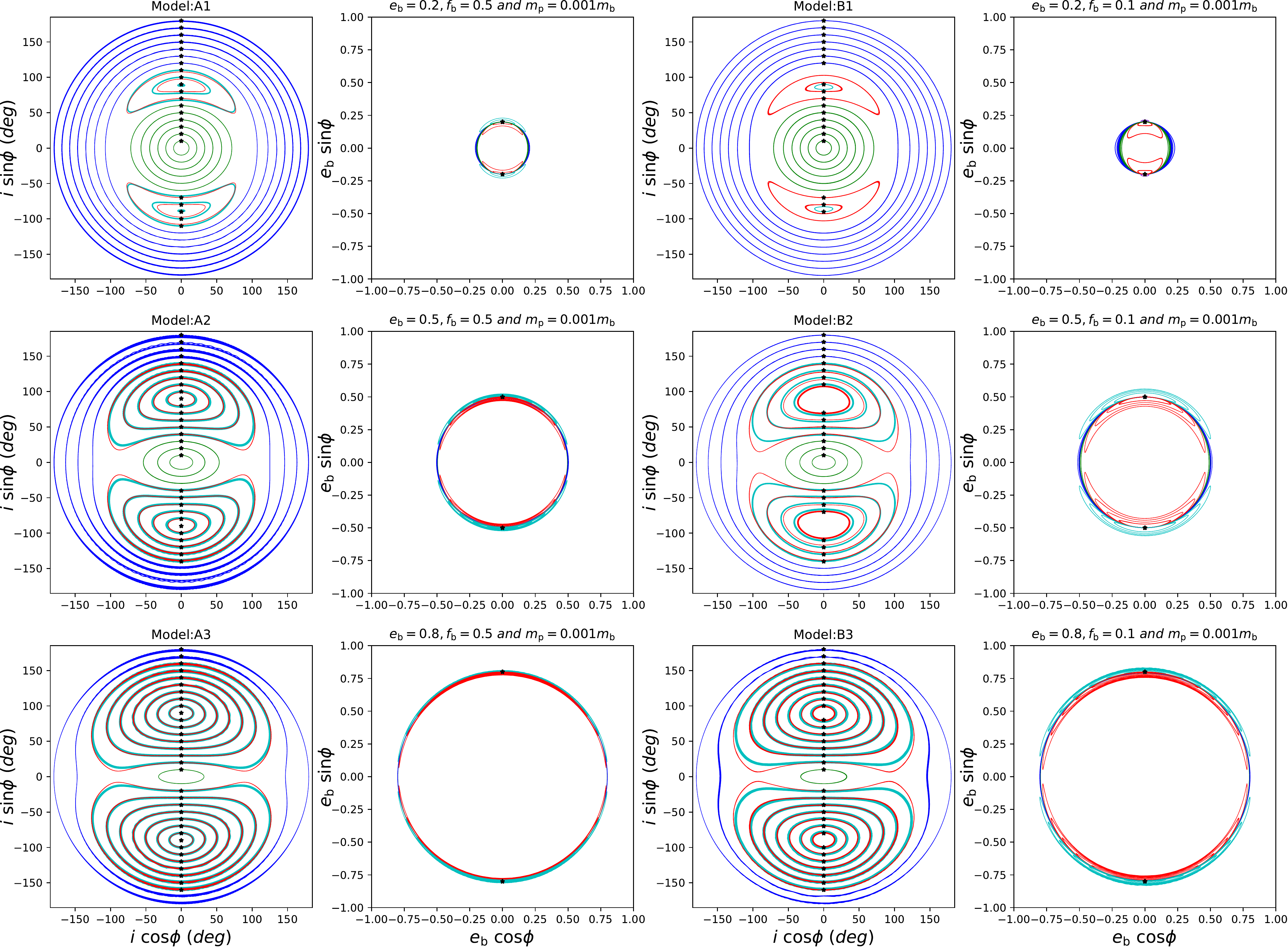}
    \caption{The $i\cos \phi-i \sin \phi$ plane (first and third columns) and $e_{\rm b}\cos{\phi}-e_{\rm b}\sin{\phi}$ plane (second and fourth columns) for orbits with different values of initial inclination and longitude of the ascending node. The planet has mass $m_{\rm p} = 0.001\, m_{\rm b}$, and orbital radius $5\,a_{\rm b}$. The binary eccentricity is $e_{\rm b}=0.2,\,0.5\,,0.8$ in the upper, middle and lower panels respectively. The mass fraction of the binary is $f_{\rm b}=0.5$ in the first and second column and $f_{\rm b}=0.1$ in the third and fourth column. The green and blue lines represent prograde and retrograde circulating orbits, respectively. The red lines represent librating orbits with initial inclination $i<i_{\rm s}$ while the cyan lines represent librating orbits with initial inclination $i>i_{\rm s}$.  The black stars mark the initial positions of the planet with $i$ ranging from 10$^\circ$ to 180$^\circ$. We removed unstable orbits.  }
    %{\bf Not sure what you mean by removing overlapping orbits. The red and cyan orbit overlap.} }
     \label{fig:surface1}
\end{figure*}

Fig.~\ref{fig:surface1} shows the results of our three--body simulations for a low mass planet with $m_{\rm p}=0.001\,m_{\rm b}$ on an orbit with semi--major axis $r=5\,a_{\rm b}$ for varying $e_{\rm b}$ and $f_{\rm b}$. Each line in each plot corresponds to a planet orbit with different initial inclination to the binary.
The first and third columns show the $i\cos \phi$--$i \sin \phi$ plane while the second and fourth columns show the corresponding $e_{\rm b}\cos \phi$--$e_{\rm b}\sin \phi$ plane. The black stars represent the initial position of the planet. 

%{\bf I'm not sure I see all the blue stars against the blue orbits. Maybe should be black. Cyan lines are hard to see. Maybe should be thicker. \CC{changed to black.}}

The green lines represent prograde (relative to the binary) circulating orbits where the planet displays clockwise precession in the longitude of the ascending node $\phi$. The blue lines correspond to retrograde circulating orbits where the planet displays counterclockwise precession in $\phi$. The red and cyan lines identify librating orbits. The inclination at the centre of these orbits is the stationary inclination, $i_{\rm s}$. In this low mass planet case, the centres are at $i=i_{\rm s}\approx 90^\circ$ and $\phi = \pm 90^\circ$. The red lines have initial inclination $i<i_{\rm s}$ while the cyan lines have initially $i>i_{\rm s}$. These librating orbits display counterclockwise precession in $\phi$. 
 
The $i\cos \phi$--$i \sin \phi$ phase plots are very similar to those in the test particle case considered by \cite{Doolin2011}. 
The eccentricity of the binary does not vary during the precession of a test particle because the particle does not have angular momentum to exchange with the binary system. 
 The $e_{\rm b}\cos \phi$--$e_{\rm b}\sin \phi$ panels for these models 
shows the curves are sometimes circulating while slightly noncircular and sometimes librating due to the eccentricity oscillations associated with the relatively small but nonzero particle (planet) mass.
 If instead the object orbiting the binary is a planet (a nonzero mass object), $e_{\rm b}$ initially increases or decreases depending on whether the initial angle of the planet's orbit is greater or smaller 
 than the stationary inclination. 
As we shall see, the larger the planet angular momentum, the larger the oscillations of the binary
orbit.

\begin{figure*}
  \centering
    \includegraphics[width=9cm]{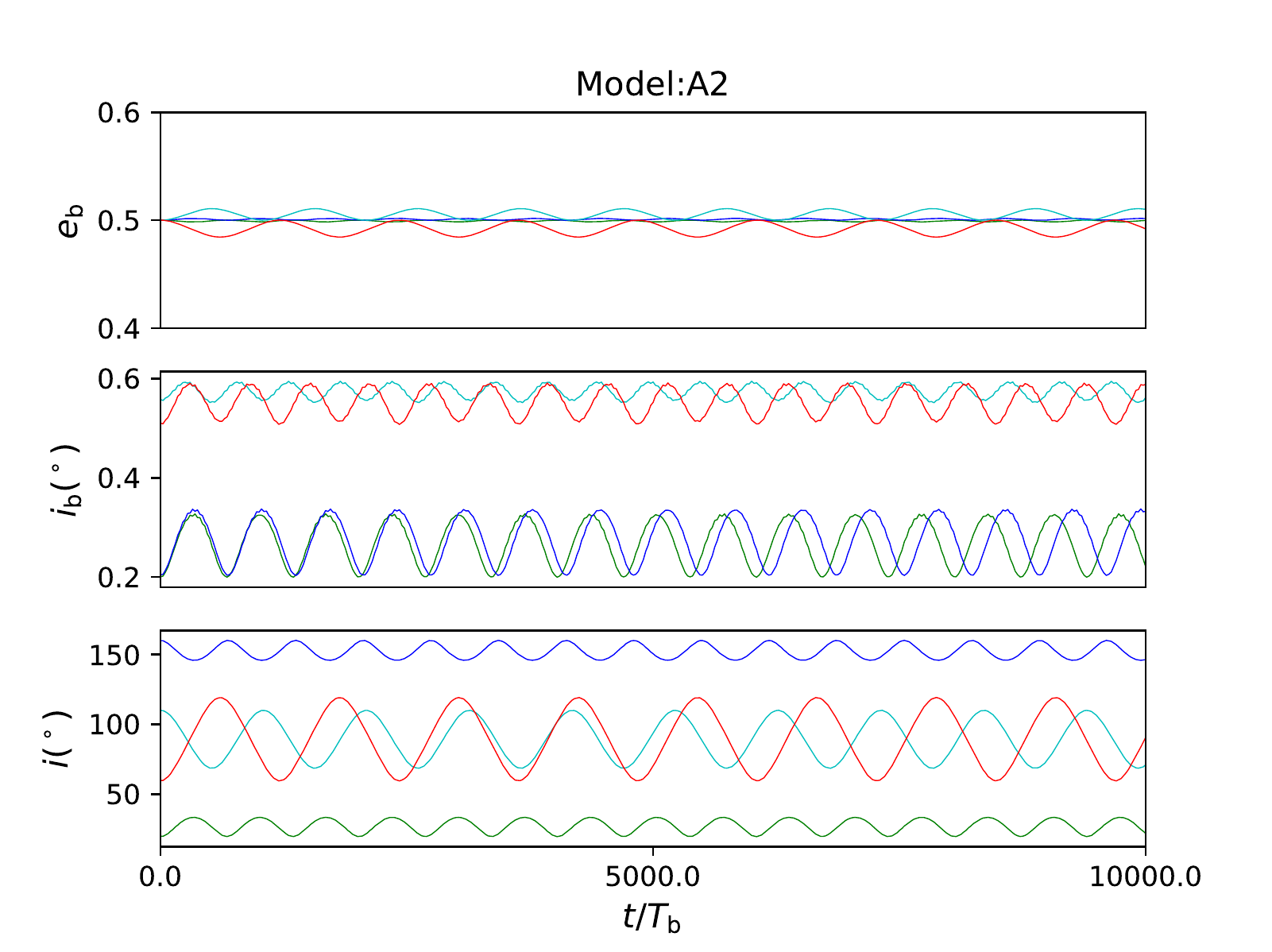}
    \hspace{-0.7cm}
    \includegraphics[width=9cm]{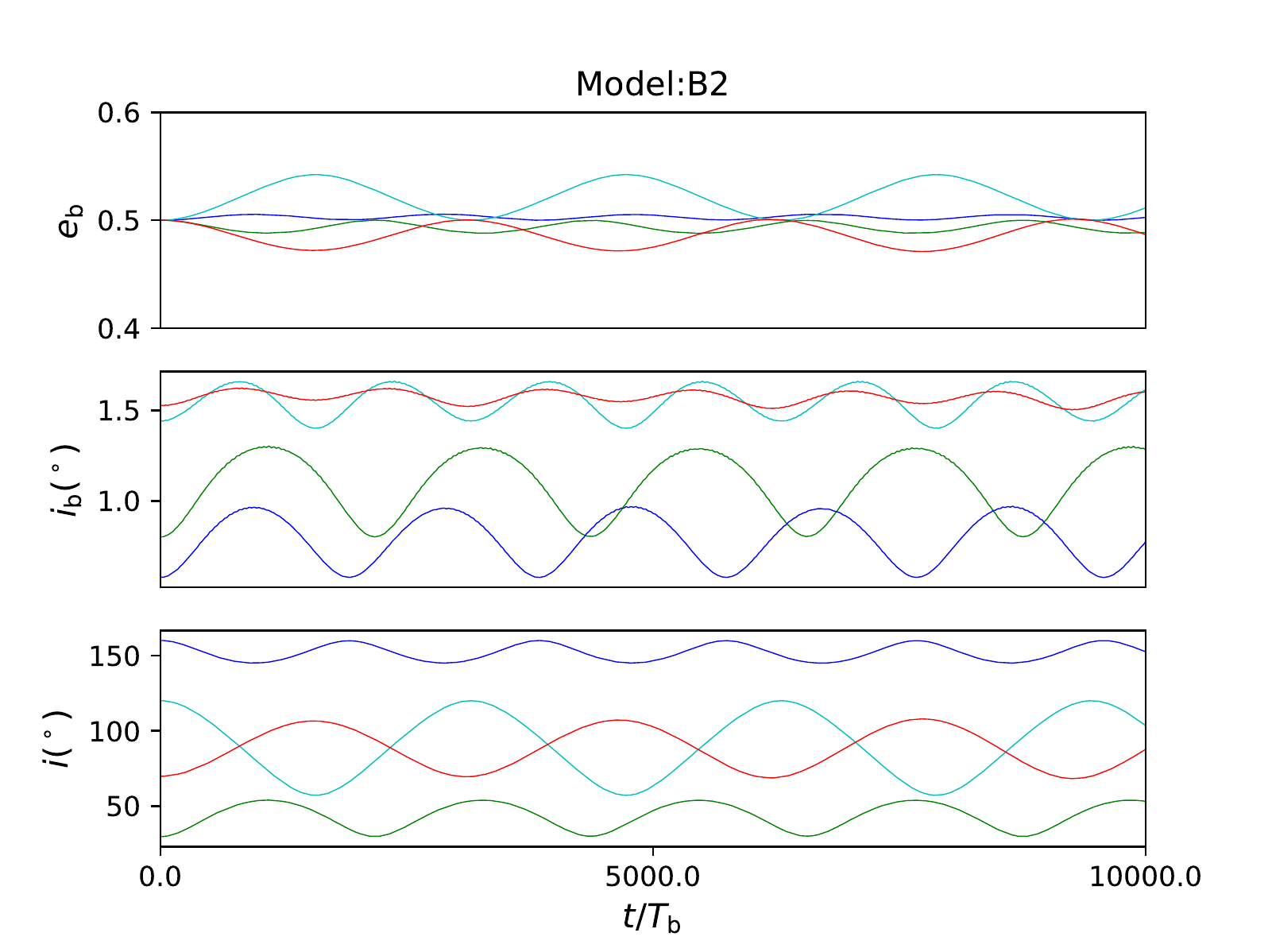}
    \includegraphics[width=9cm]{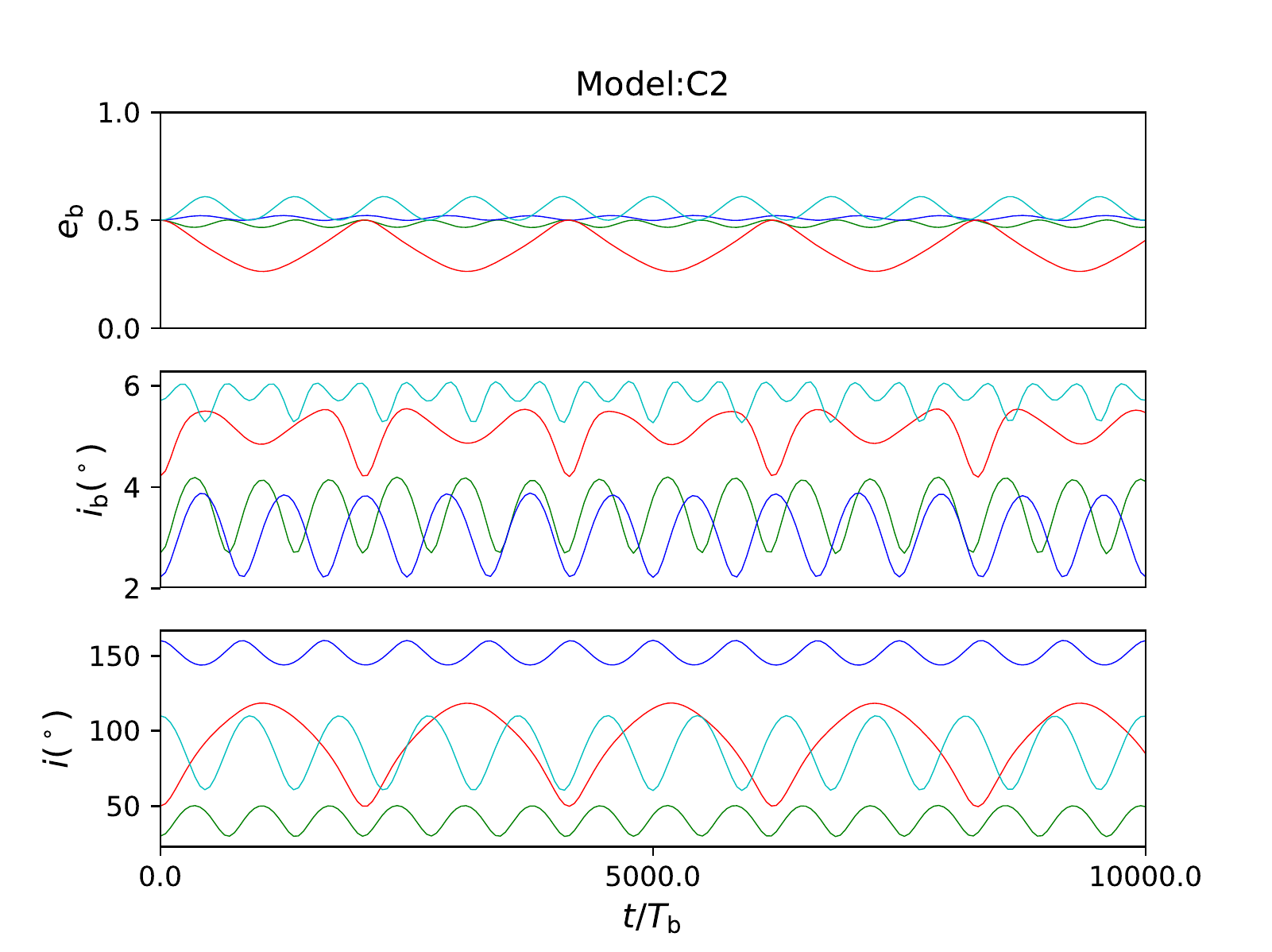}
    \hspace{-0.7cm}
    \includegraphics[width=9cm]{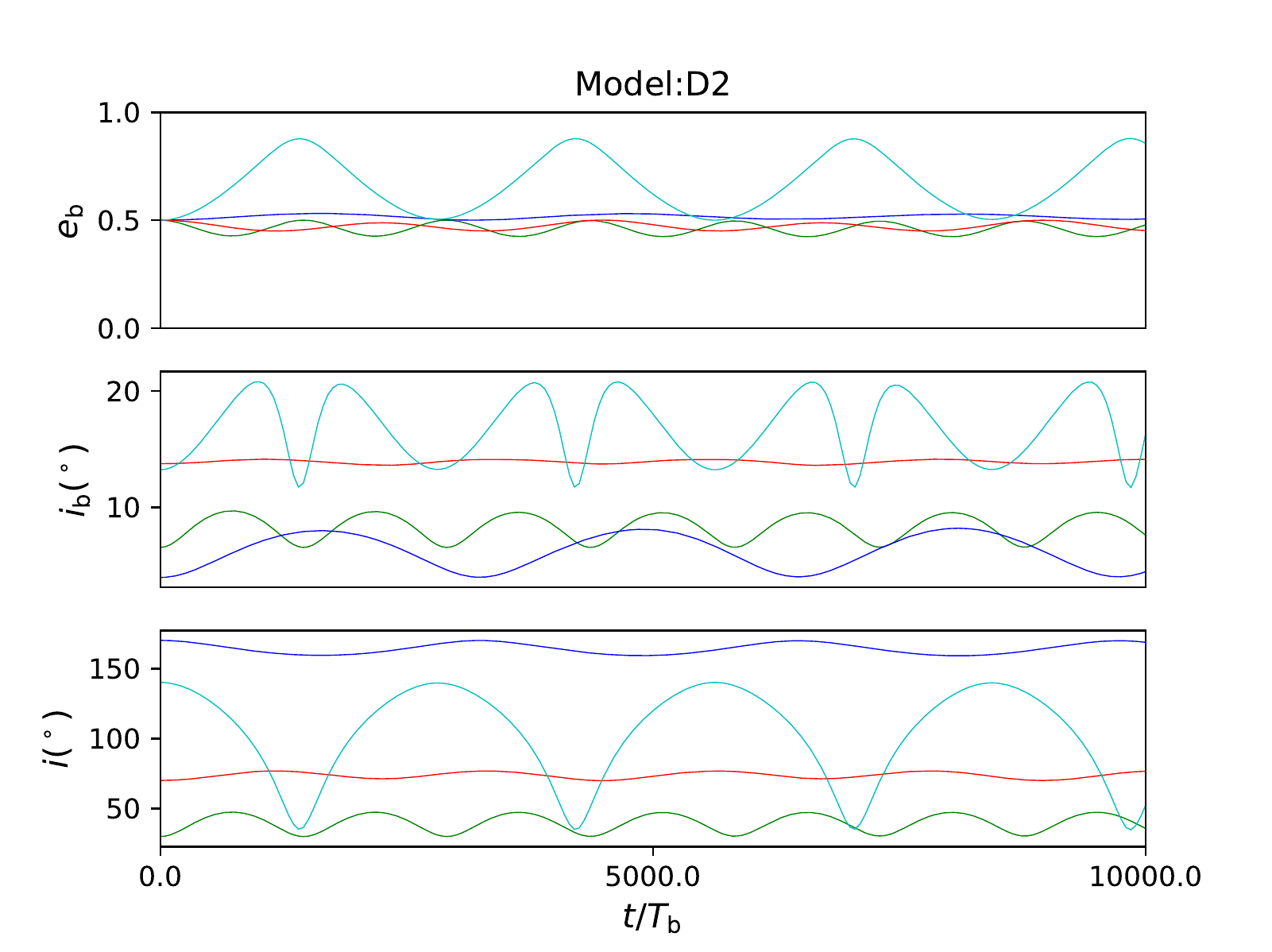}
    \includegraphics[width=9cm]{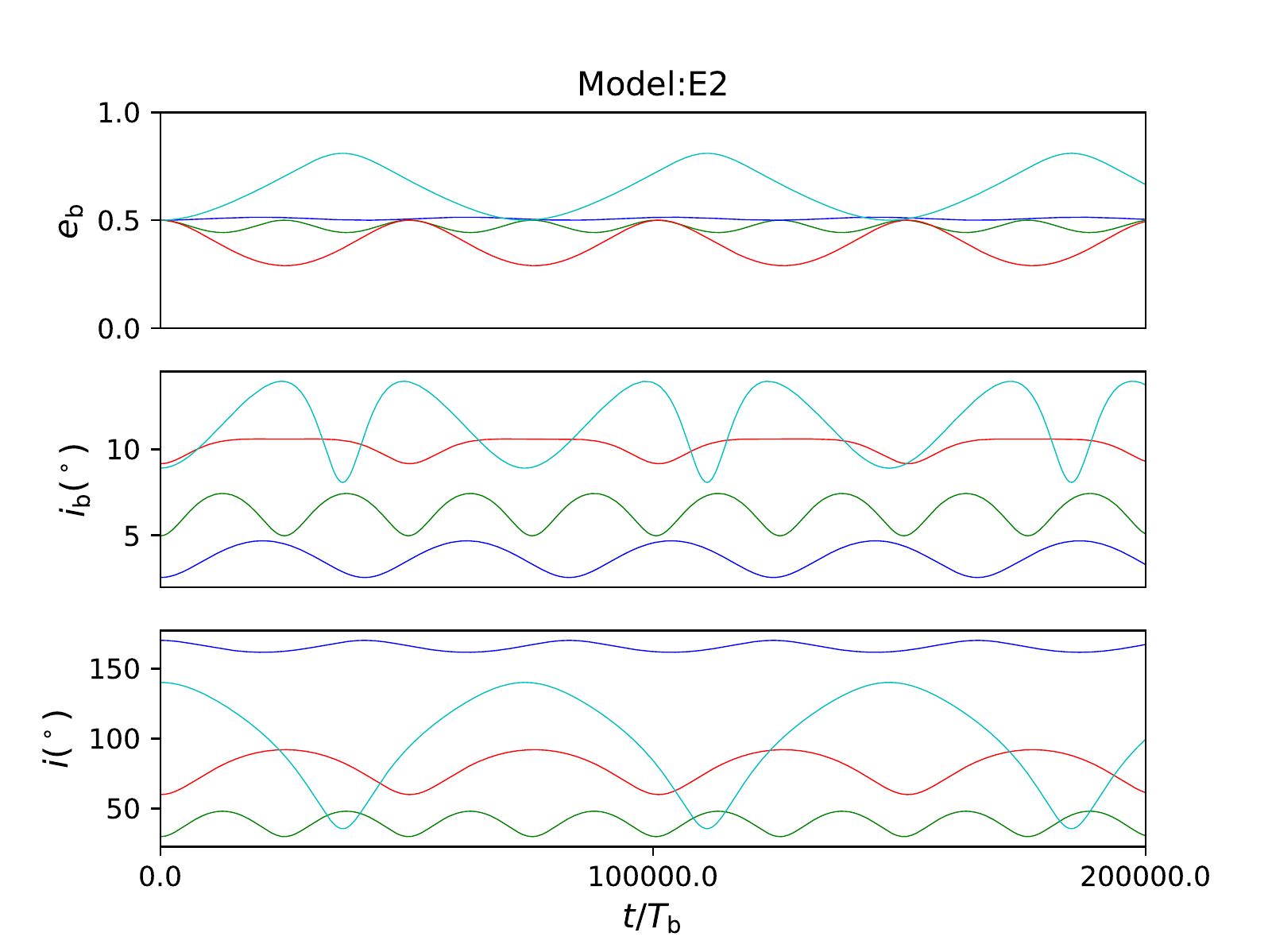}
    \hspace{-0.65cm}
    \includegraphics[width=9cm]{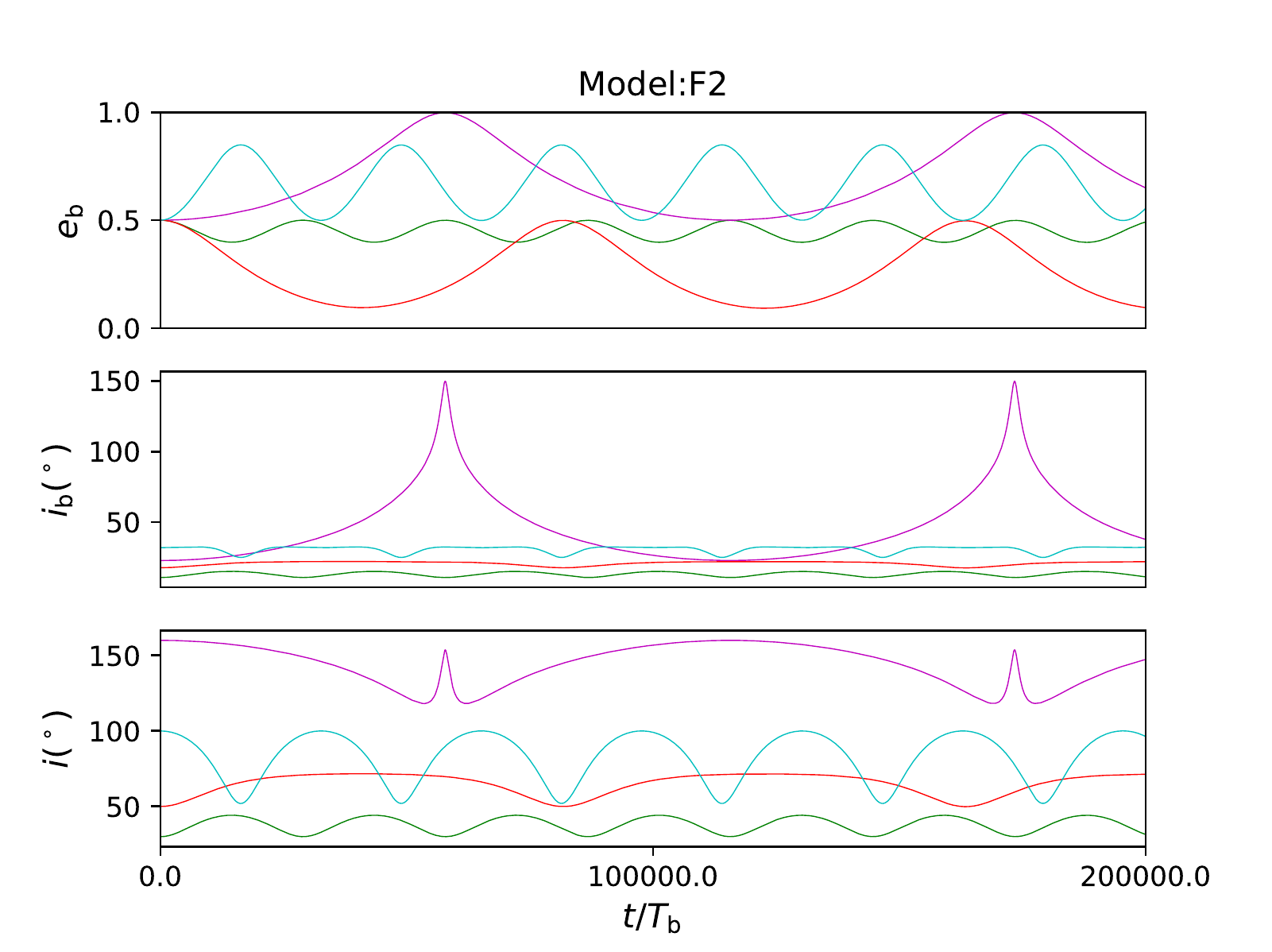}
    \caption{Time evolution of the binary eccentricity, $e_{\rm b}$, the inclination of the binary $i_{\rm b}$ with respect to the vector of the total angular momentum, and the inclination of the planet  with respect to the vector of the binary angular momentum,$i$, for Models A2, B2, C2 D2, E2 and F2. Each panel contains one line for each different type of orbit with the difference being the initial inclination which is shown in the bottom panel. The colours correspond to the orbit type described in Fig.~\ref{fig:surface1}.}
    \label{fig:surfaceinfo}
\end{figure*}

%\subsubsection{Effect of the binary eccentricity}

In order to investigate the effect of the initial binary eccentricity, we ran simulations with initial $e_{\rm b}$ = 0.2, 0.5 and 0.8. Comparing the three rows in Fig.~\ref{fig:surface1}, we see that higher initial eccentricities correspond to larger libration islands. 
%Since the planet mass is small, the binary eccentricity $e_{\rm b}$ does not vary significantly during the planet precession in the $e_{\rm b}\cos \phi$--$e_{\rm b}\sin \phi$ plane because the angular momentum of the binary dominates the system.

%\subsubsection{Effect of the binary mass fraction}

The comparison between Models A and B in Fig.~\ref{fig:surface1} shows the effect of decreasing the binary mass fraction from $f_{\rm b}=0.5$ to $f_{\rm b}=0.1$. 
The size and shape of the orbits in the $i\cos \phi$--$i \sin \phi$ plane does not change significantly with mass fraction, but the variations of $e_{\rm b}$ in the libration region are larger for smaller binary mass fraction.
We find that for lower binary mass fraction, the orbits near the libration centre (with inclination $i=i_{\rm s}$ and phase angle $\phi=\pm90^\circ$) become divergent for moderate to high binary eccentricities ($e_{\rm b}$ = 0.5, 0.8) and therefore might be unstable.  This issue is beyond the scope of this work. We will explore the stability of orbits close to misaligned binaries in a future publication. 

In the top row of Fig.~\ref{fig:surfaceinfo} we consider in more detail the evolution of the orbits in time for models with initial binary eccentricity of 0.5 and binary mass fraction $f_{\rm b}=0.5$ (top left, model A2) and $f_{\rm b}=0.1$ (top right, model B2). We show $e_{\rm b}$, $i_{\rm b}$, and $i$  as a function of time for different values of the initial planet inclination. The colour of the lines corresponds to the orbit type in Fig.~\ref{fig:surface1}. Comparing the left and right plots, we see that there is less 
evolution of the binary in the equal mass binary compared to the lower binary mass fraction. This is because with lower binary mass fraction, the planet has more angular momentum compared to the binary and thus has a stronger effect on the binary. The timescale for the oscillations is longer for smaller binary mass fraction.

\subsection{High mass planet at small orbital radius}

\begin{figure*}
  \flushleft
    \includegraphics[width=17.7cm]{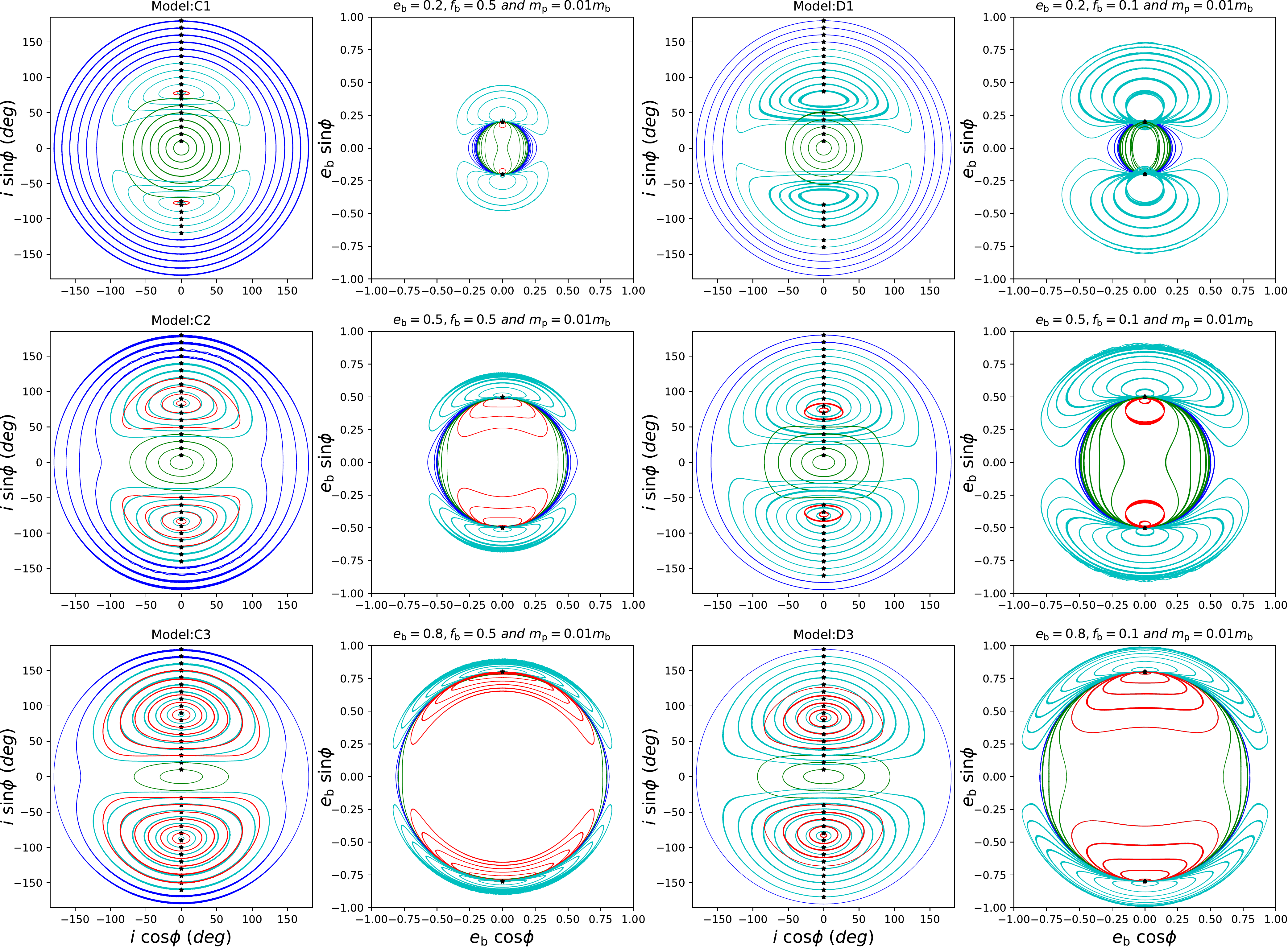}
    \caption{Same as Fig.~\ref{fig:surface1} except $m_{\rm p} = 0.01\,m_{\rm b}$.
}
     \label{fig:surface2}
\end{figure*}

Fig.~\ref{fig:surface2} shows the results in the $i\cos \phi$--$i \sin \phi$ and $e_{\rm b}\cos \phi$--$e_{\rm b}\sin \phi$ plane obtained with a higher mass planet  $m_{\rm p} = 0.01\,m_{\rm b}$ orbiting an eccentric binary at $r=5\,a_{\rm b}$. The line colours correspond to the same type of orbits as in Fig.~\ref{fig:surface1}. 
Comparing Fig.~\ref{fig:surface2} with Fig.~\ref{fig:surface1}, we see that the region of prograde circulating orbits is  larger for a higher mass planet. The differences between the red and cyan lines becomes more prominent. The inclination at the centre of the libration, $i_{\rm s}$, decreases. For example, for model C2 with $e_{\rm b}=0.5$ and $f_{\rm b}=0.5$ it is 80$^\circ$. Because of this decrease, there is a narrower range of inclinations for which red orbits exist, those with initial inclination $i<i_{\rm s}$, and a wider range of inclinations for which cyan orbits exist with $i>i_{\rm s}$ initially. We discuss the value of the stationary inclination further in Section~\ref{critical}.

The higher planet mass causes the binary eccentricity to vary significantly  not only in the librating solutions, but also in the circulating orbits of the planet. The binary eccentricity in the libration islands for Model  C2 in Fig.~\ref{fig:surface2} starts at $e_{\rm b}=0.5$ and reaches values as large as $e_{\rm b} \simeq 0.7$.

By comparing Models C and D in Fig.~\ref{fig:surface2}, we can see the effect of changing the binary mass fraction. The cyan libration islands are significantly larger and there are fewer red orbits in the simulation with $f_{\rm b}$ = 0.1 because $i_{\rm s}$ is smaller than in the equal mass binary case. 
Therefore, decreasing the binary mass fraction results in a lower stationary inclination. We find particle orbital instability close to $i=i_{\rm s}$ and $\phi=90^\circ$ for Model D1.  There are no stable orbits in the region close to the stationary inclination.

The variations in $e_{\rm b}$ are larger for the smaller binary mass fraction, as seen in the $e_{\rm b}\cos \phi$--$e_{\rm b}\sin \phi$ plot. In particular, $e_{\rm b}$ in Model D3 becomes very close to 1 during libration.

The middle panels of Fig.~\ref{fig:surfaceinfo} shows the binary eccentricity, inclination and planet inclination evolution with time for Models C2 and D2. 
We see again that decreasing the binary mass fraction leads to larger amplitude oscillations in the binary eccentricity for librating orbits starting above the critical angle (cyan line).
Model D2 has larger variations of $i_{\rm b}$ and $i$ during the libration of the planet because the system has a larger planet to binary angular momentum ratio and the secondary star ($m_2 =0.1 m_{\rm b}$) interacts more strongly with the planet.

\subsection{High mass planet at large orbital radius}
\label{highandlarge}

\begin{figure*}
  \centering
    \includegraphics[width=17.7cm]{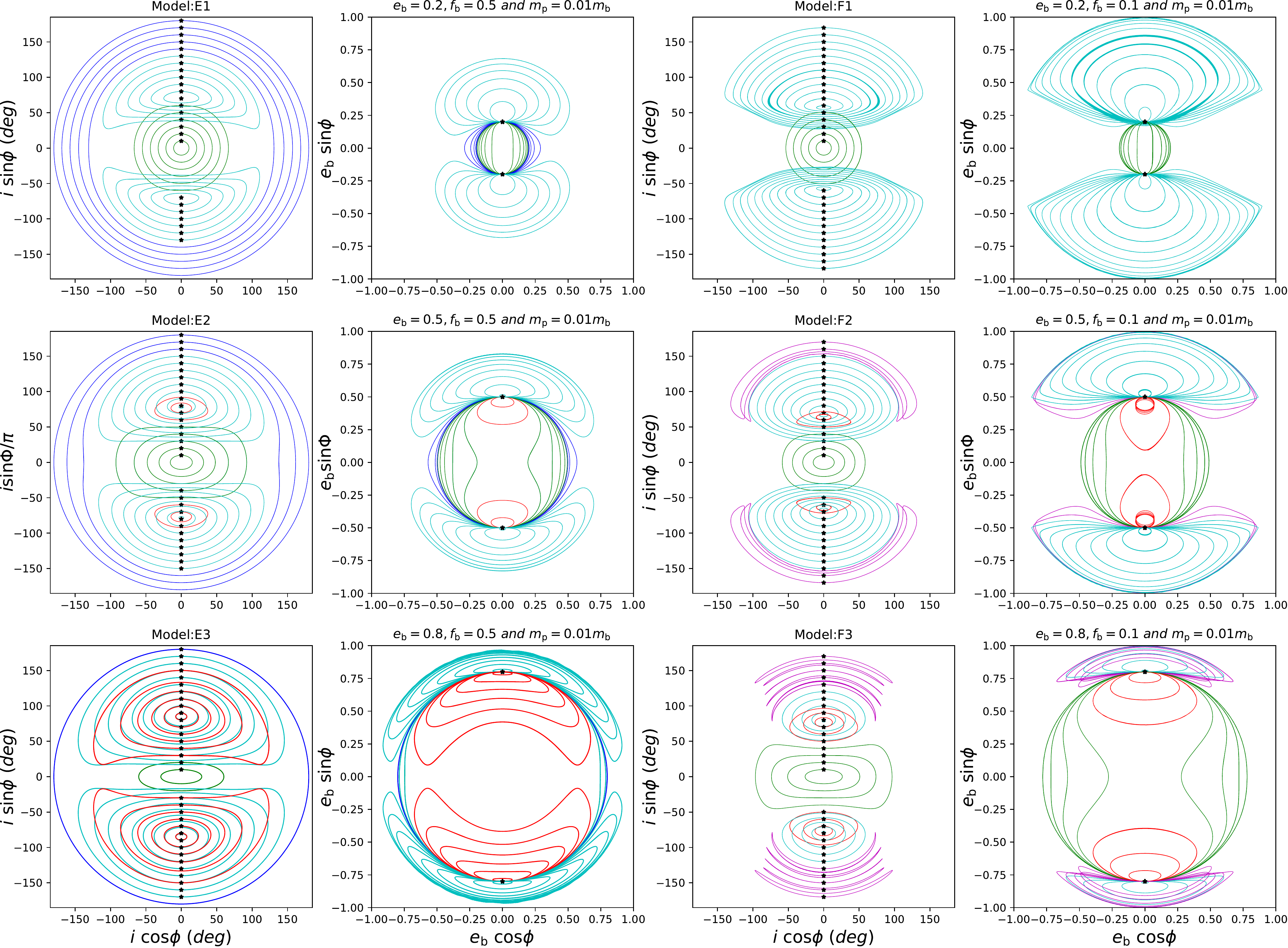}
    \caption{Same as Fig.~\ref{fig:surface2} except the orbital radius is 20 $a_{\rm b}$. The magenta lines show crescent orbits. }
     \label{fig:surface3}
\end{figure*}

We now increase the planet semi--major axis to $r=20\,a_{\rm b}$ and keep its mass at $m_{\rm p}=0.01\,m_{\rm b}$. Comparing Models E in Fig.~\ref{fig:surface3} with Models C in Fig.~\ref{fig:surface2}, we see that the libration islands become even larger when the planet orbits the binary with a larger semi-major axis. The eccentricity of the binary $e_{\rm b}$ can be excited to larger values during the planet's libration because in this configuration the planet has more angular momentum to exchange with the binary system. 
The increase in the initial binary eccentricity $e_{\rm b}$ has the same effect as in previously described simulations, i.e., it increases the  range of stable librating orbits.

We find that there are no retrograde circulating solutions for the low mass fraction case $f_{\rm b}=0.1$ (see the third column of Fig.~\ref{fig:surface3}, there are no blue lines). 
%Equation A1 of \cite{MartinandLubow2019} {\bf There should always be retrograde circulating orbits with $i=180^\circ$, coplanar retrograde. Can you verify that? } 
The eccentricity of the binary can be excited to very large values (close to 1) during the planet's precession in Models F1, F2, and F3.  However, there is a new type of orbit that is different from those described in previous sections. These librating orbits, represented by magenta lines, only appear in the simulations with smaller binary mass fraction $f_{\rm b}=0.1$ that start with higher initial eccentricities $e_{\rm b}=0.5,\,0.8$. The magenta orbits have higher initial inclination than the librating cyan lines  and display counterclockwise precession. In Model F2, these crescent shape orbits appear in the $i\cos \phi$--$i \sin \phi$ plane for $i > \ 150^{\circ}$ while in Model F3, they appear also for lower inclinations $i> 140^{\circ}$. These librating orbits are not nested within each other, as they are in the prograde case.

%  Libration centres occur at stationary tilts between the planet and binary. 
Appendix A of \cite{MartinandLubow2019} shows that noncoplanar stationary states for retrograde 
orbits with $i_{\rm s} < 180^\circ$ do not exist below a critical value of the angular momentum ratio $j_{\rm cr}$ given by
equation A3 in that paper 
\begin{equation}
    j_{\rm cr}= \frac{1+4 e_{\rm b}^2}{2+3 e_{\rm b}^2}.
    \label{jcrit}
\end{equation}
The lack of stationary states means that the crescent shaped librating orbits are always of non-zero extent in the $i\cos \phi --i\sin \phi$ phase plane.
This is understood by the lack of stationary retrograde inclinations for $j < j_{\rm cr}$. For example, for Model F2 $j=0.58 < j_{\rm cr}=1.29$ and for Model F3 $j=0.83 < j_{\rm cr}=0.91$.

Appendix A of \cite{MartinandLubow2019} shows analytically
that stationary coplanar retrograde $i=180^\circ$ (and also prograde $i=0^\circ$) orbits should exist.
However, we have not been able to find such an orbit numerically for Models F1, F2, and F3. 

In the bottom panels of Fig.~\ref{fig:surfaceinfo}, we show the evolution of $e_{\rm b}$, $i_{\rm b}$, and $i$ for Model E2 and F2. 
The time oscillations of $e_{\rm b}$, $i_{\rm b}$ and $i$ have longer periods compared to the corresponding case of the same planet mass with smaller semi-major axis. 
We show the evolution of one of the crescent orbits, the magenta lines in Model F2. As $e_{\rm b}$ is very close to 1 during the precession, the vector of the binary angular momentum changes quickly, resulting in the narrow peaks in the $i_{\rm b}$ and $i$ plots.  

\subsection{Systems with high angular momentum ratios }
\label{highangular}

To investigate the retrograde librating orbits, we now consider simulations of an equal mass binary with higher angular momentum ratios $j = 1$ and $j=2$, with binary eccentricities $e_{\rm b}$ = 0.2, 0.5, and 0.8.  The angular momentum ratios are larger than the critical required for a retrograde libration centre given in  Equation~(\ref{jcrit}). 

\begin{figure*}
  \flushleft
    \includegraphics[width=17.7cm]{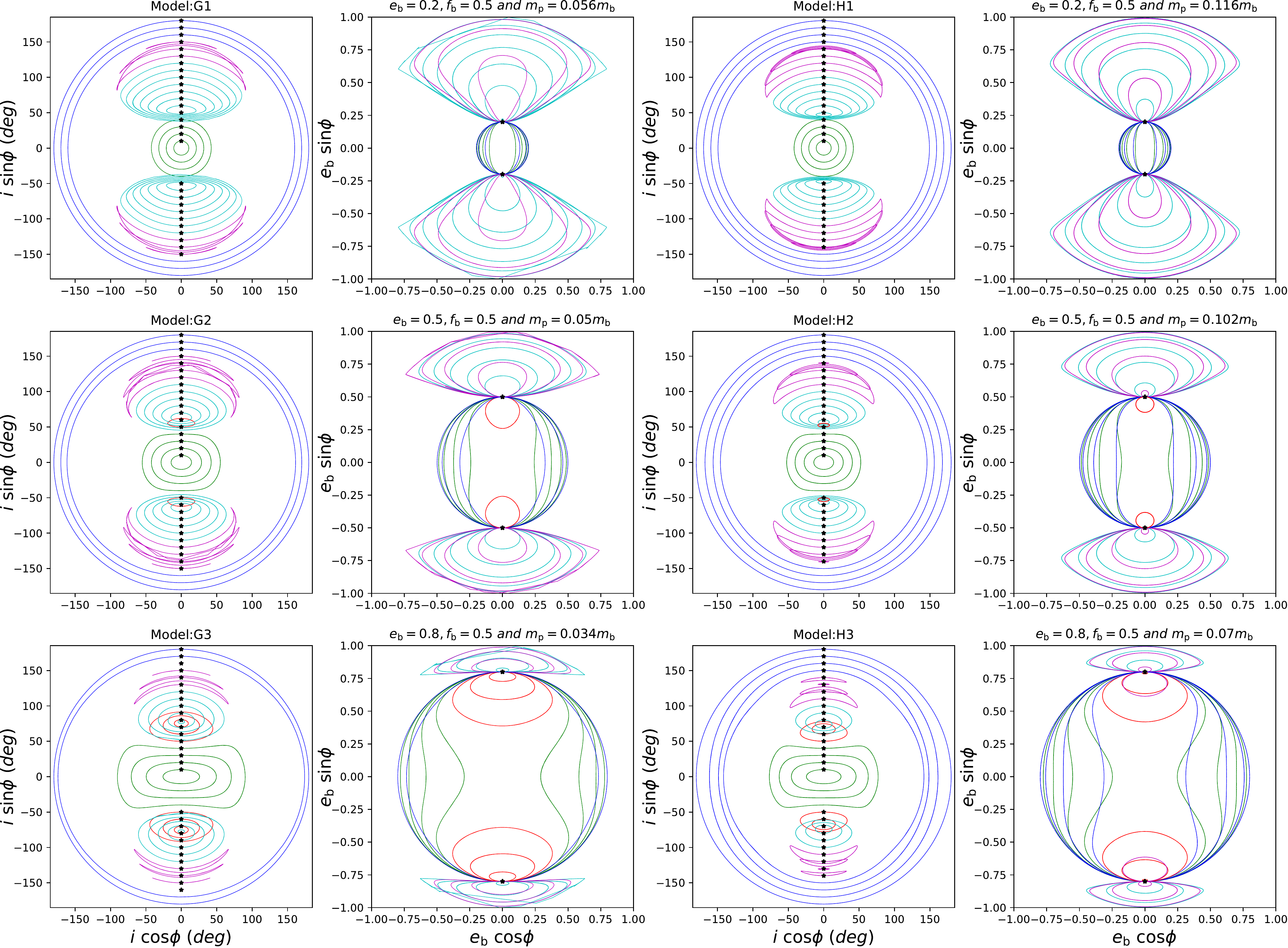}
    \caption{Same as Fig.~\ref{fig:surface3} except that each model has different $m_{\rm p}$ to satisfy $j=1$ (left) and $j=2$ (right).}
     \label{fig:surface4}
\end{figure*}

Fig.~\ref{fig:surface4} shows the orbital evolution of  simulations with $j = 1$ (left panels) and $j=2$ (right panels). 
There are  librating orbits in the phase diagrams in the first and third columns of Fig.~\ref{fig:surface4} that surround the stationary points. with $i_{\rm s} < 90 ^\circ$. The fully retrograde librating orbits ($i > 90^ \circ$ throughout the orbit) are seen as the crescent magenta orbits in the $i\cos \phi -i\sin \phi$ phase plane.  In the case of inclinations less than the critical inclination, the orbits  decrease in extent in the phase plane with increasing initial inclination. They are at most  only partially overlapping and are not fully nested. They reach zero extent at the stationary angle, $i_{\rm s}$. For inclinations above the stationary angle, the orbits increase in extent in the phase plane. 
Above the critical angle, the binary eccentricity initially decreases while below the critical angle the binary eccentricity initially increases starting at $\phi=90^\circ$, indicated by the stars in the phase planes. Unlike the prograde librating case, the retrograde librating orbits orbits are not nested about a common centre that occurs at the stationary inclination.

\section{Stationary and critical angles}
\label{critical}
 
We compare the results of our numerical simulations to the analytic results presented in \cite{MartinandLubow2019}  for the stationary inclination (inclination at the centre of the libration island), $i_{\rm s}$, and the critical minimum inclination angle $i_{\rm min}$ that separates the prograde orbits from the librating orbits. We also calculate numerically the critical maximum inclination for librating orbits, $i_{\rm max}$.  The analytic results are based on secular equations with the quadrupole approximation for the binary potential \citep{Farago2010}. The equations are expected to break down for orbits that are close to the binary.

\subsection{Stationary inclination}

The stationary inclination $i_{\rm s}$ depends only the eccentricity of the binary, $e_{\rm b}$, and the ratio of the angular momentum of the particle to the angular momentum of the binary, $j$.  The equation is given by 
\begin{equation}
\cos i_{\rm s} = \frac{-(1+4e_{\rm b}^2) \pm \sqrt{(1+4e_{\rm b}^2)^2+60(1-e_{\rm b})j^2}}{10j}\,
\label{eq:ic}
\end{equation}
\citep[see equation 17 in][]{MartinandLubow2019}, where we take the positive square root for prograde orbits ($\cos i_{\rm s}>0)$ and the negative square root for retrograde orbits ($\cos i_{\rm s} <0$; see Appendix A of \cite{MartinandLubow2019}).

\subsubsection{Prograde Stationary Inclination}

\begin{figure}
    \centering
    \includegraphics[width=8.8cm]{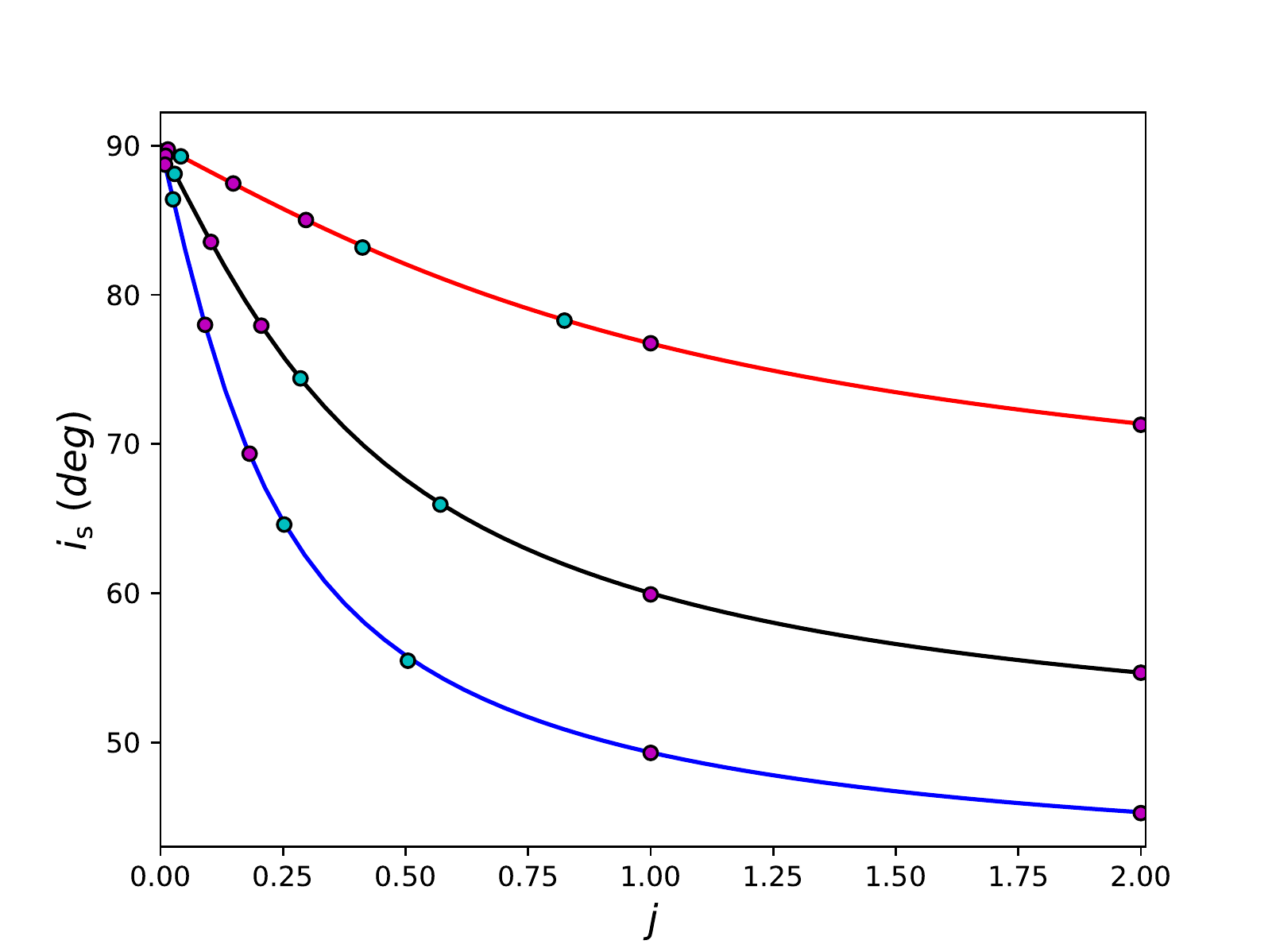}
    \caption{Comparison of the analytic solution given by Equation (\ref{eq:ic}) in the prograde case (solid lines) with simulation results (dots) for the stationary tilt $i_{\rm s}$ of the planet relative to the binary as a function of planet-to-binary angular momentum ratio $j$.  The solid curves have binary eccentricities $e_{\rm b}$ = 0.2 (blue line), 0.5 (black line), and 0.8 (red line). Cyan dots correspond to models with $f_{\rm b}$ = 0.1 and magenta dots correspond to models with $f_{\rm b}$ = 0.5. The upper set of dots is for simulations of models that have $e_{\rm b} = 0.8$, the middle set have $e_{\rm b} = 0.5$, and the lower set have $e_{\rm b} = 0.2$.}
    \label{fig:ac}
\end{figure}

The solid lines in Fig.~\ref{fig:ac} plot Equation (\ref{eq:ic}) in the prograde case for $e_{\rm b}$ = 0.2 (blue line), 0.5 (black line) and 0.8 (red line) as a function of the planet-to-binary angular momentum ratio $j$.  For fixed $e_{\rm b}$, the stationary inclination $i_{\rm s}$ decreases monotonically with increasing $j$ and for fixed $j$ it increases monotonically with increasing $e_{\rm b}$.  The stationary inclinations for the models in Table \ref{table1} were determined numerically from the simulations. The magenta dots in Fig.~\ref{fig:ac}  correspond to the models with $f_{\rm b}=0.5$ and the cyan dots correspond to the models with $f_{\rm b}=0.1$. The results confirm the prediction that for fixed binary eccentricity the stationary inclination depends directly on $j$, independent of $f_{\rm b}$.  
The simulations are in very good agreement with the analytically results.  
 
\begin{figure} 
    \centering
    \includegraphics[width=8.8cm]{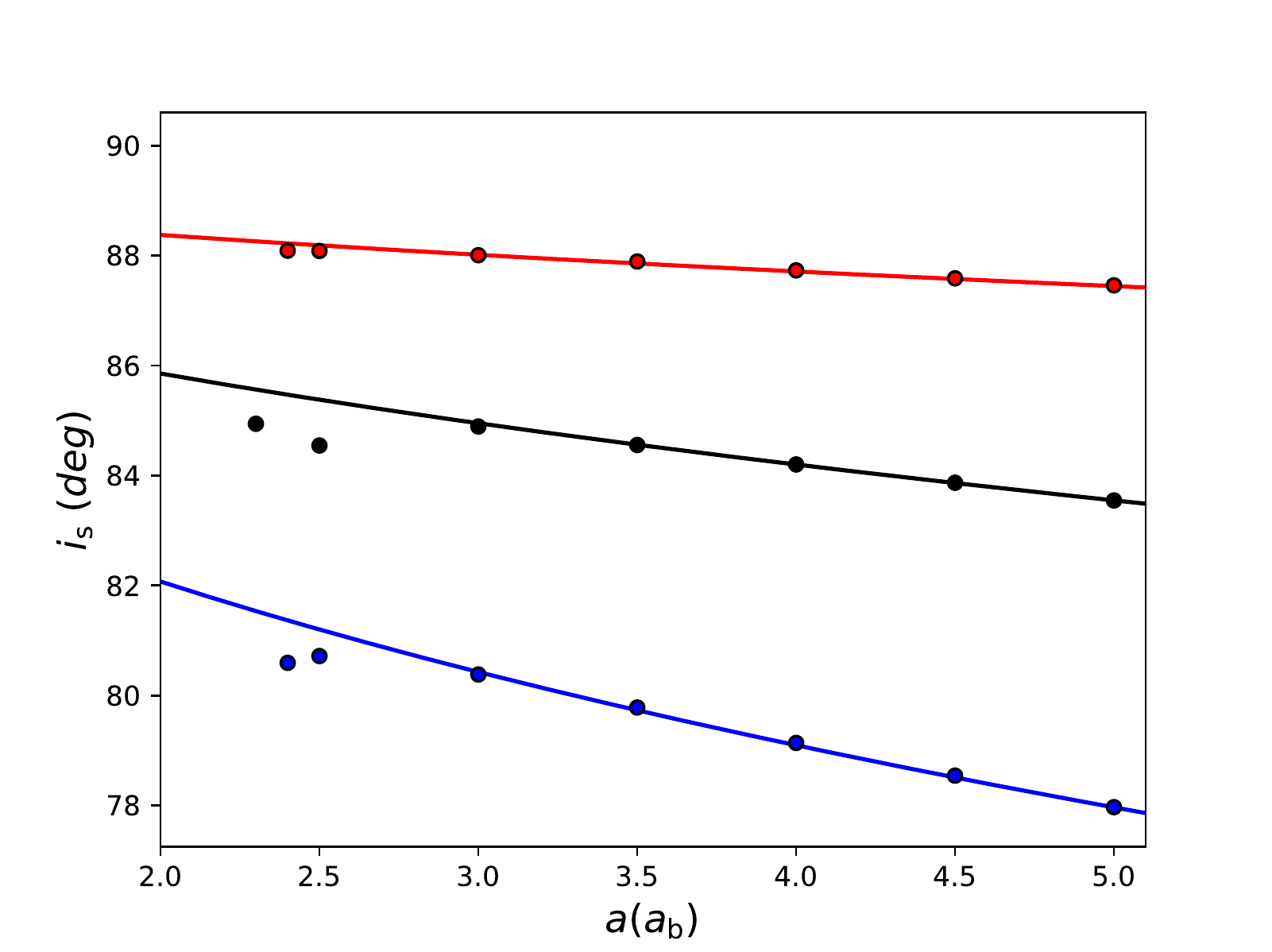}
    \caption{Comparison of the analytic solution given by Equation (\ref{eq:ic}) in the prograde case (solid lines) with simulation results (dots) for the stationary tilt $i_{\rm s}$ of the planet relative to the binary as a function of the semi-major axis of the planet $a$. 
    The solid lines have $e_{\rm b}$ = 0.2 (blue line), 0.5 (black line), and 0.8 (red line). The upper set of dots is for simulations of models that have $e_{\rm b} = 0.8$, the middle set have $e_{\rm b} = 0.5$, and the lower set have $e_{\rm b} = 0.2$.}
    \label{fig:inner}
\end{figure}

 The quadrupole approximation made in deriving Equation (\ref{eq:ic})  is more accurate for larger planet orbital radii $a$. 
To test the accuracy of the prograde analytic solution, we consider simulations with a close-in planet. In Fig.~\ref{fig:inner} we plot Equation (\ref{eq:ic}) in the prograde case for $e_{\rm b}$ = 0.2 (blue line), 0.5 (black line), and 0.8 (red line) as a function of $a$. The points show simulation results for numerically determined stationary inclinations. The dot colours correspond to the binary eccentricity of the analytic lines. The orbit of the planet is unstable if $a$ is less than about $2.3 a_{\rm b}$. Thus, the innermost dots which we acquire by our simulations are for a planet at $a$ = 2.3 (black line) and 2.4 (red and blue lines). The simulations are in very good agreement with the analytically results at large $a$, but deviate somewhat at smaller values of  $a$. 

The black dotted lines in Fig.~\ref{fig:critical1} and Fig.~\ref{fig:critical2} plot the prograde stationary inclinations $i_{\rm s}$ from our simulations and the magenta lines in the same figures plot the analytic solution for the prograde stationary inclination given by Equation (\ref{eq:ic}) as a function of binary eccentricity $e_{\rm b}$ for 
 all parameters fixed except the binary eccentricity. Fig.~\ref{fig:critical1} shows the results for a planet at $r=5\,a_{\rm b}$ while Fig.~\ref{fig:critical2} refers to the same system but with the planet at $r=20\,a_{\rm b}$.  The upper panels and bottom panels of two figures show the critical angles for an equal mass binary and a binary with $f_{\rm b}=0.1$ respectively. The left and right panels of the two figures show results for different planet masses. The black dotted lines are in very good agreement with the analytic results. 
The prograde stationary inclination  values for the low mass planet are rather insensitive to the location of the planet or $f_{\rm b}$ and so the lines look similar and are in the range of $80^\circ-90^\circ$, since $j$ is small (see Fig.~\ref{fig:ac}). On the other hand, the stationary inclination is sensitive to $e_{\rm b}$ for the high mass planet.  The stationary inclination angle is smaller for smaller binary eccentricity and smaller binary mass fraction
(larger $j$).

\subsubsection{Retrograde Stationary Inclination}

%In Sections~\ref{highandlarge} and~\ref{highangular}, we showed that the retrograde stationary inclination $i_{\rm rs}$ occurs at the inner edge of the retrograde orbit  the $i\cos \phi$--$i \sin \phi$ plane, since no tilt or phase oscillations occur at there. For retrograde orbits, the equation is given by 
%\begin{equation}
%\cos i_{\rm s} = \frac{-(1+4e_{\rm b}^2)-\sqrt{(1+4e_{\rm %\label{eq:irc}
%\end{equation}
%\citep[see equation 17 in][]{MartinandLubow2019}. 

The retrograde stationary inclination is given by Equation (\ref{eq:ic}), where we take the negative square root.
The solid lines in Fig.~\ref{fig:irc} show the analytical solutions for $e_{\rm b}$ = 0.2 (blue line), 0.5 (black line), and 0.8 (red line) as a function of angular momentum ratio $j$. The retrograde stationary inclination, $i_{\rm s}$, decreases monotonically with increasing $j$. However, the behaviour with binary eccentricity is more complicated, as we discuss further below. The six dots whose colours correspond to the $e_{\rm b}$ values for the analytic curves that represent $i_{\rm s}$ of Models G1 to H3. The simulations are in very good agreement with the analytic predictions.
 
There are two major qualitative differences between the prograde and retrograde stationary orbits.
In the prograde case, for any value of planet-to-binary angular momentum ratio $j$ there is a stationary
inclination value about which there are nested librating orbits in the $i \cos \phi$-- $i \sin \phi$
plane (e.g., Figs.~\ref{fig:surface1} and \ref{fig:inner}). But in the retograde case, we see from 
Fig.~\ref{fig:irc} that $i_{\rm s}$ reaches $180^\circ$ for $j=j_{\rm cr}$ given by Equation (\ref{jcrit}).
For $j < j_{\rm cr}$ there are no stationary noncoplanar librating orbits, as discussed
in Section \ref{highandlarge}. 

The second difference between the prograde and retrograde stationary orbits is that for any fixed $j$, the stationary inclination angle
$i_{\rm s}$ increases monotonically with $e_{\rm b}$ in the prograde case, but not generally in the retrograde case.
The difference is seen by comparing Fig.~\ref{fig:inner} (prograde) and Fig.~\ref{fig:irc} (retrograde) in that the curves
in the prograde case for different $e_{\rm b}$ do not intersect, except at $j=0$ and $i_{\rm s}=90^\circ$  that is the upper limit of prograde tilts, while in the retrograde case they do intersect and cross. 

To confirm this crossing in the retrograde case, recognise that the intersection implies 
that $i_{\rm s}$ is independent of $e_{\rm b}$ at some fixed value of $j$.
This condition can be written as
\begin{equation}
    \frac{\partial \cos{i_{\rm s}}}{\partial e_{\rm b}} \Big |_j=0,
\end{equation}
where $\cos{i_{\rm s}}$ is given by Equation (\ref{eq:ic}) in the retrograde case.
This condition has an analytic solution for the intersection point at $j=j_{\rm int}$ and $i_{\rm s}=i_{\rm int}$,
where
\begin{eqnarray}
j_{\rm int} &=& \frac{2}{\sqrt{3}}, \\  
i_{\rm int} &=& 150^\circ,
\end{eqnarray}
which is in agreement with the point of intersection in Fig.~\ref{fig:irc}.
For fixed $j$ greater (smaller) than $j_{\rm int}$, the stationary angle $i_{\rm s}$ increases with decreasing (increasing) 
binary eccentricity $e_{\rm b}$.

\subsection{Critical inclinations for libration}
 
The critical minimum inclination angle between the prograde circulating and librating orbits can be determined analytically. We follow the description in Section~3.4 of \cite{MartinandLubow2019}. There are two branches, one with lower $j$ and one with higher $j$, based on the the sign of the parameter $\chi$ that is defined as
\begin{equation}
\chi = e_{\rm b}^2-2(1-e_{\rm b}^2)j(2j+\cos i)\,.
\label{chi}
\end{equation}
The minimum tilt angle occurs where $\phi=90^\circ$.
If $\chi$ >0, the minimum tilt angle for libration to occur is
\begin{equation}
\cos i_{\rm min} = \frac{\sqrt{5}e_{\rm b}\sqrt{4e_{\rm b}^2-4j^2(1-e_{\rm b}^2)+1}-2j(1-e_{\rm b}^2)}{1+4e_{\rm b}^2},
\label{eq:icrit1}
\end{equation}
while if $\chi$<0, we have
\begin{equation}
\cos i_{\rm min} = \frac{\sqrt{(1-e_{\rm b}^2)(1+4e_{\rm b}^2+60(1-e_{\rm b}^2)j^2)}-(1-e_{\rm b}^2)}{10(1-e_{\rm b}^2)j}\,.
\label{eq:icrit2}
\end{equation}

\begin{figure*}
  \centering
    \includegraphics[width=8.7cm]{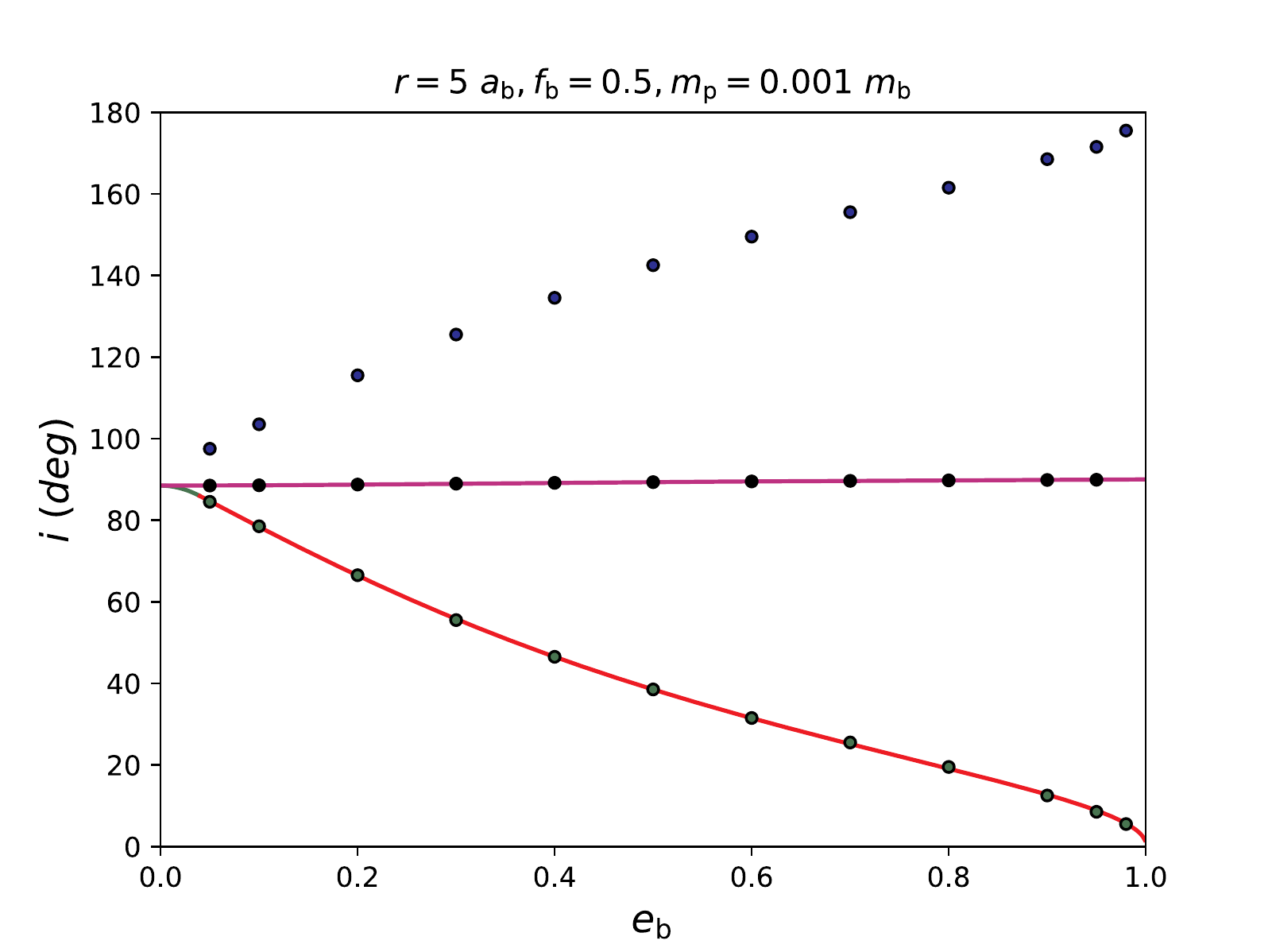}
    \includegraphics[width=8.7cm]{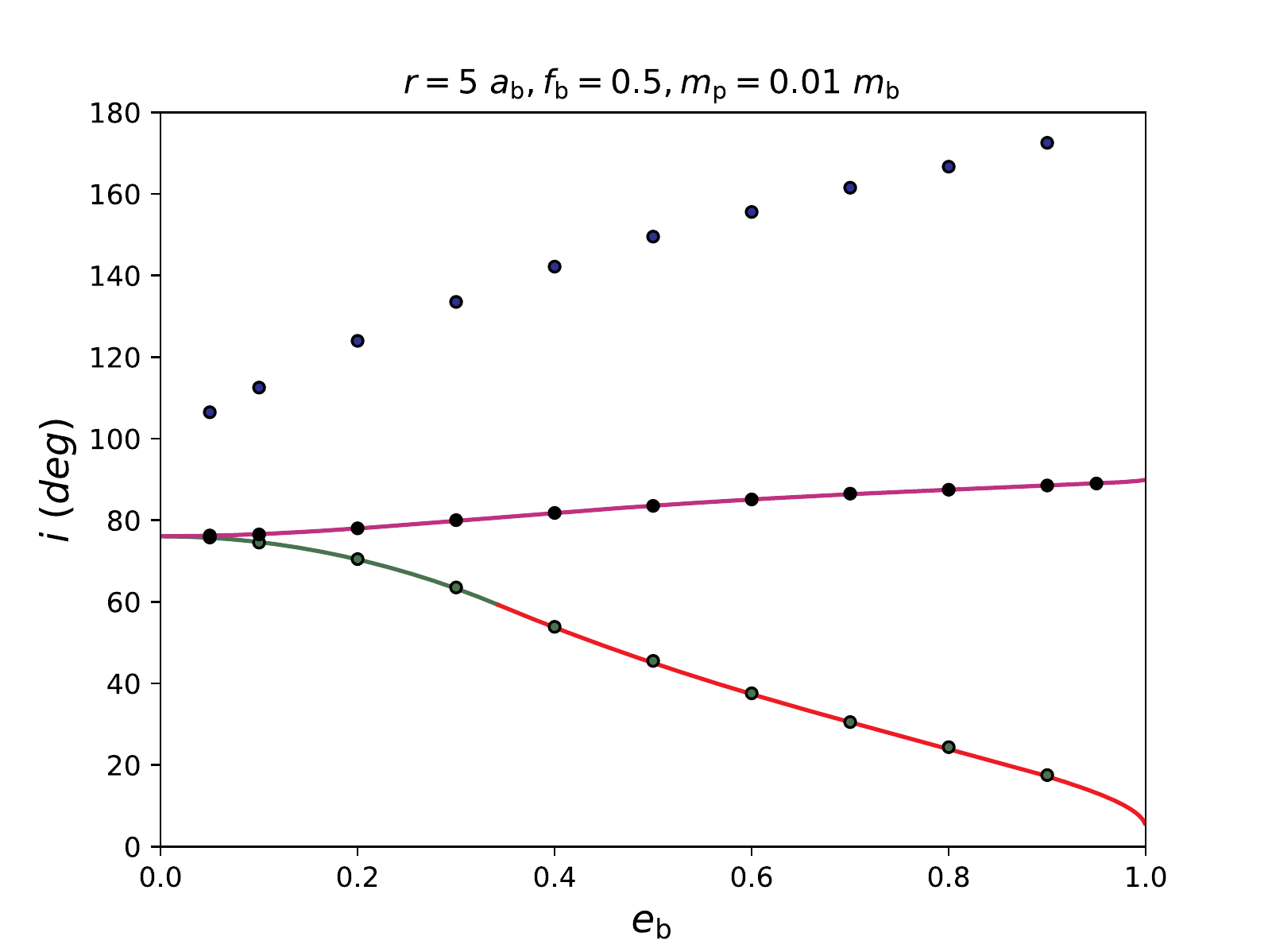}
    \includegraphics[width=8.7cm]{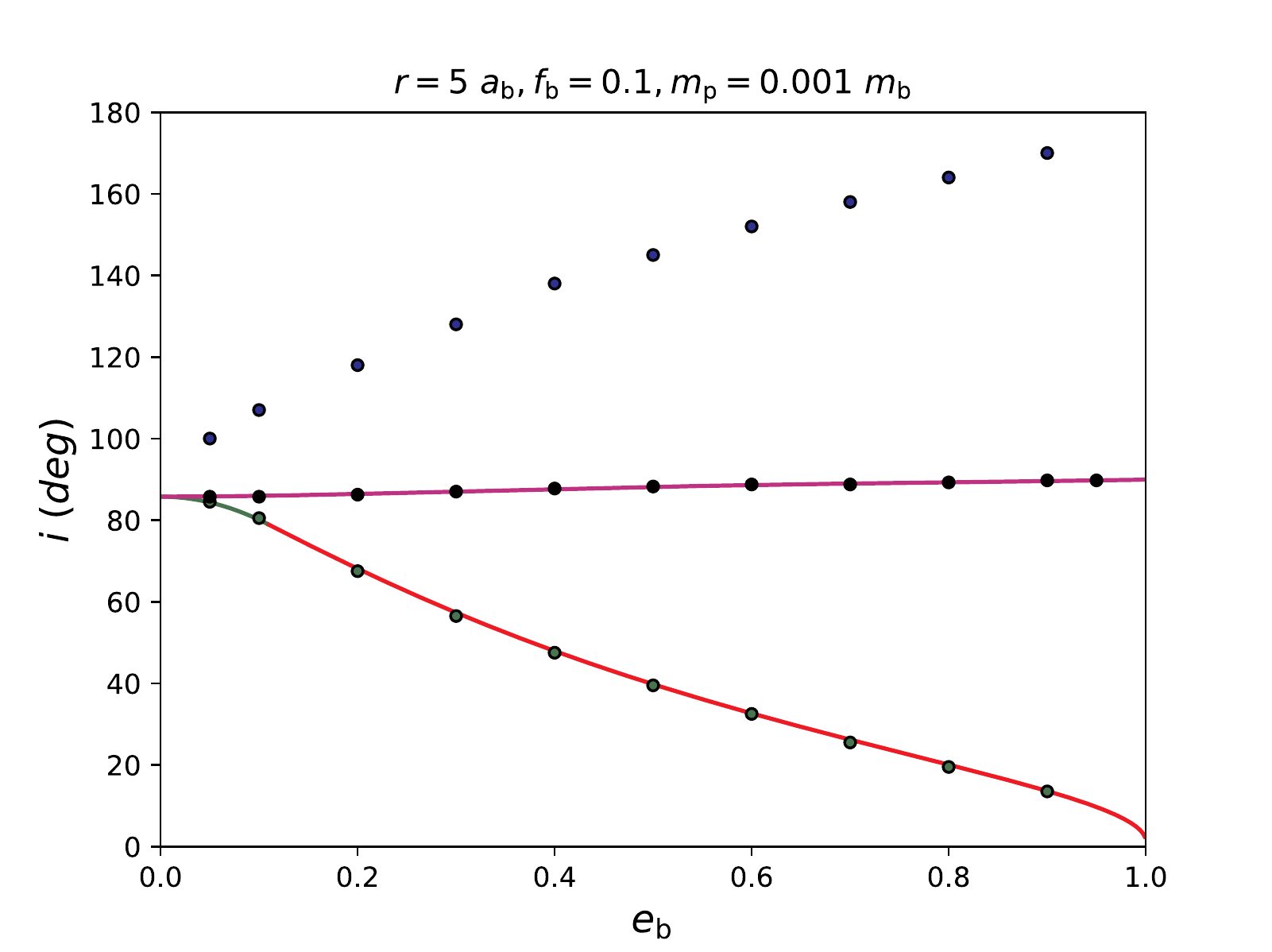}
    \includegraphics[width=8.7cm]{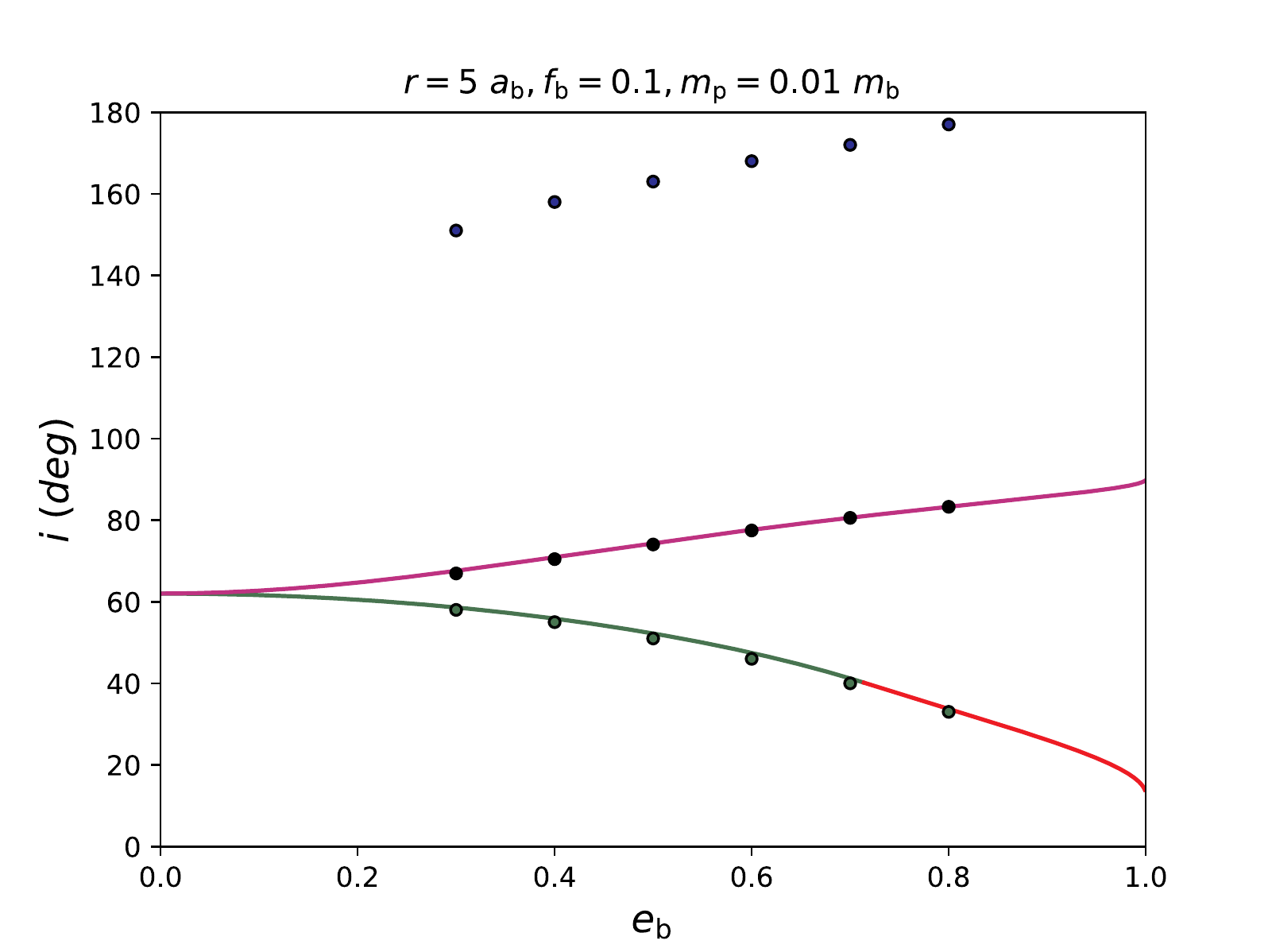}
    \caption{ Stationary inclination ($i_{\rm s}$), critical minimum inclination ($i_{\rm min}$) for libration, and critical maximal inclination for libration as a function of the binary eccentricity with the planet orbiting at $r = 5 a_{\rm b}$ with different binary mass fractions $f_{\rm b}$ =0.5 (upper panels) and 0.1 (lower panels) for the lower mass planet (left panels) and the high mass planet (right panels). The dotted lines plot the results of numerical simulations, while the solid lines are from the analytic model. The green dotted lines show the boundary between the prograde circulating and librating orbits while the blue dotted lines show the boundary between librating and retrograde circulating orbits. The black dotted lines show $i_{\rm s}$ obtained from our simulations. The magenta lines plot the analytic solutions for $i_{\rm s}$ from Equation (\ref{eq:ic}). The red lines plot the analytic solutions for $i_{\rm min}$ from Equation (\ref{eq:icrit1}) and the green solid lines plot the analytic solutions for $i_{\rm min}$ from Equation (\ref{eq:icrit2}).}
     \label{fig:critical1}
\end{figure*}

\begin{figure*}
  \centering
    \includegraphics[width=8.7cm]{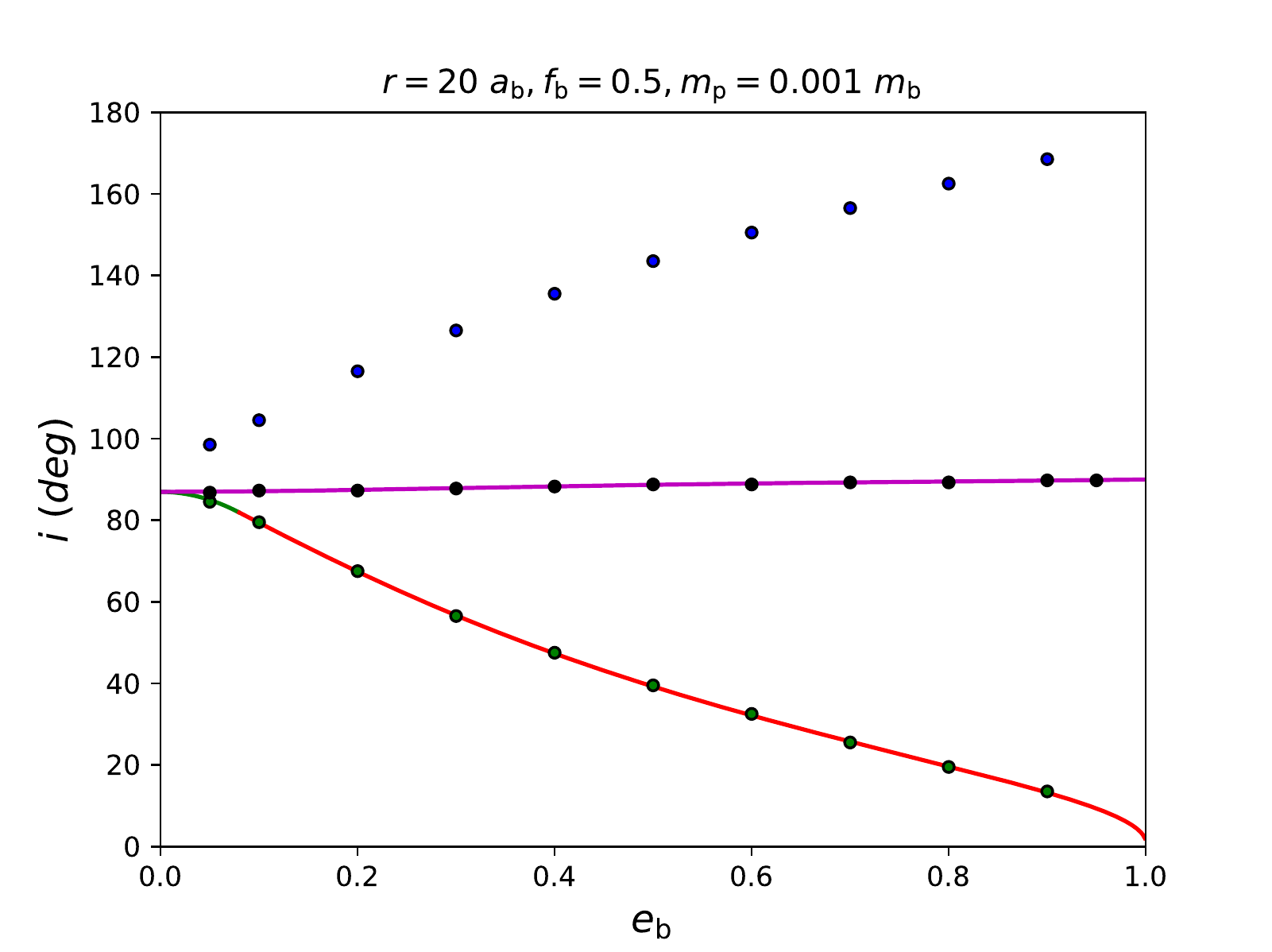}
    \includegraphics[width=8.7cm]{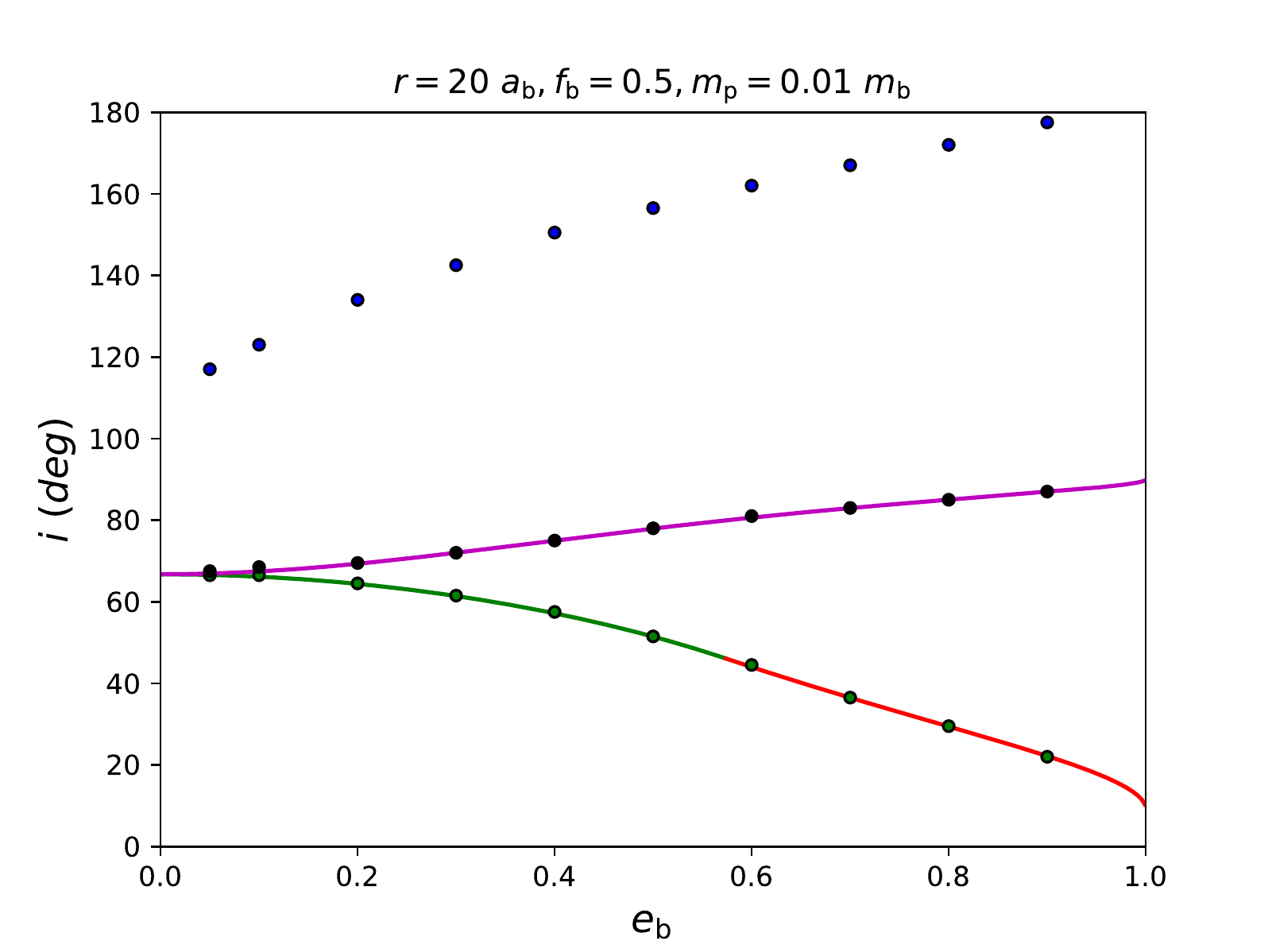}
    \includegraphics[width=8.7cm]{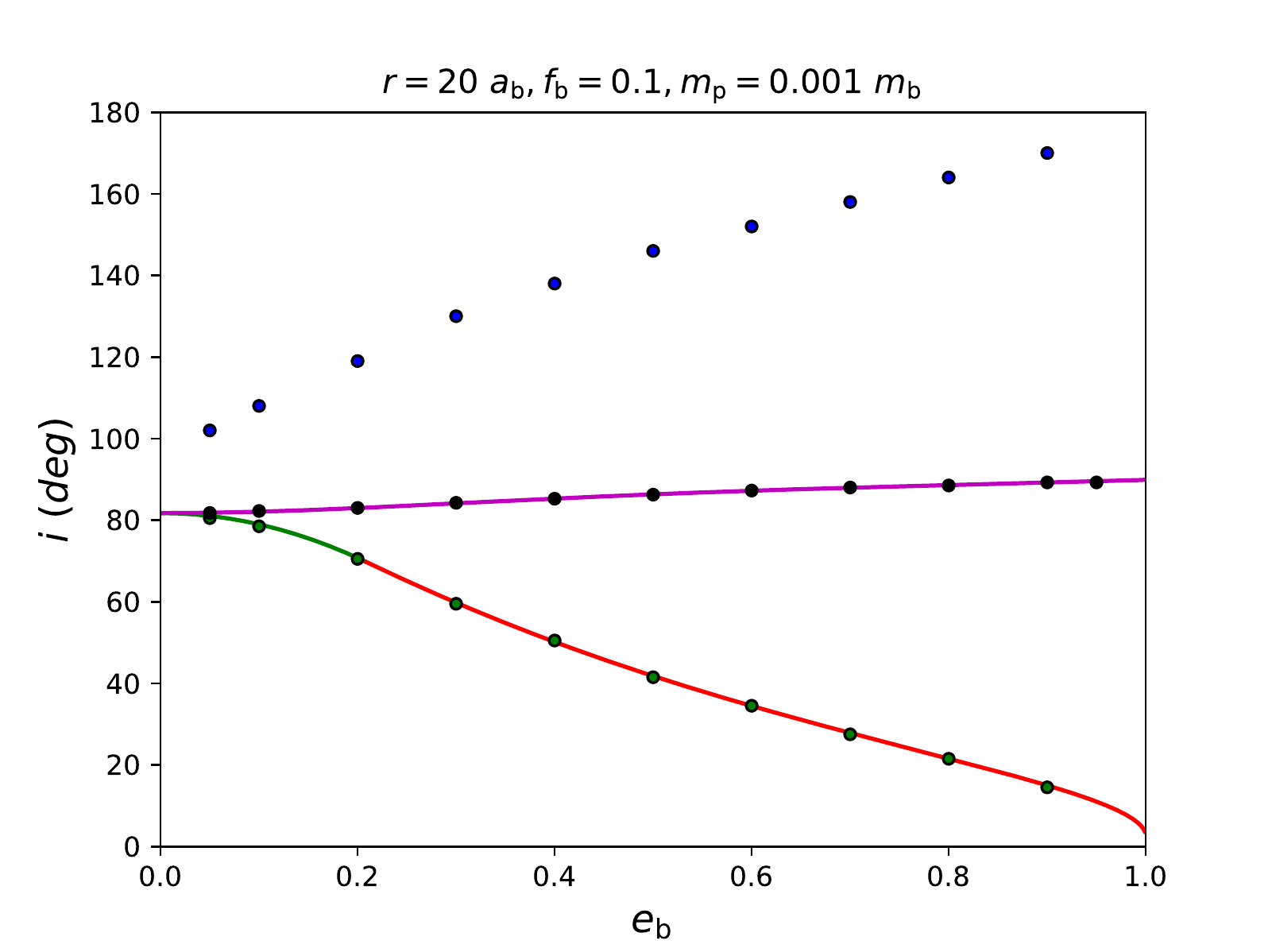}
    \includegraphics[width=8.7cm]{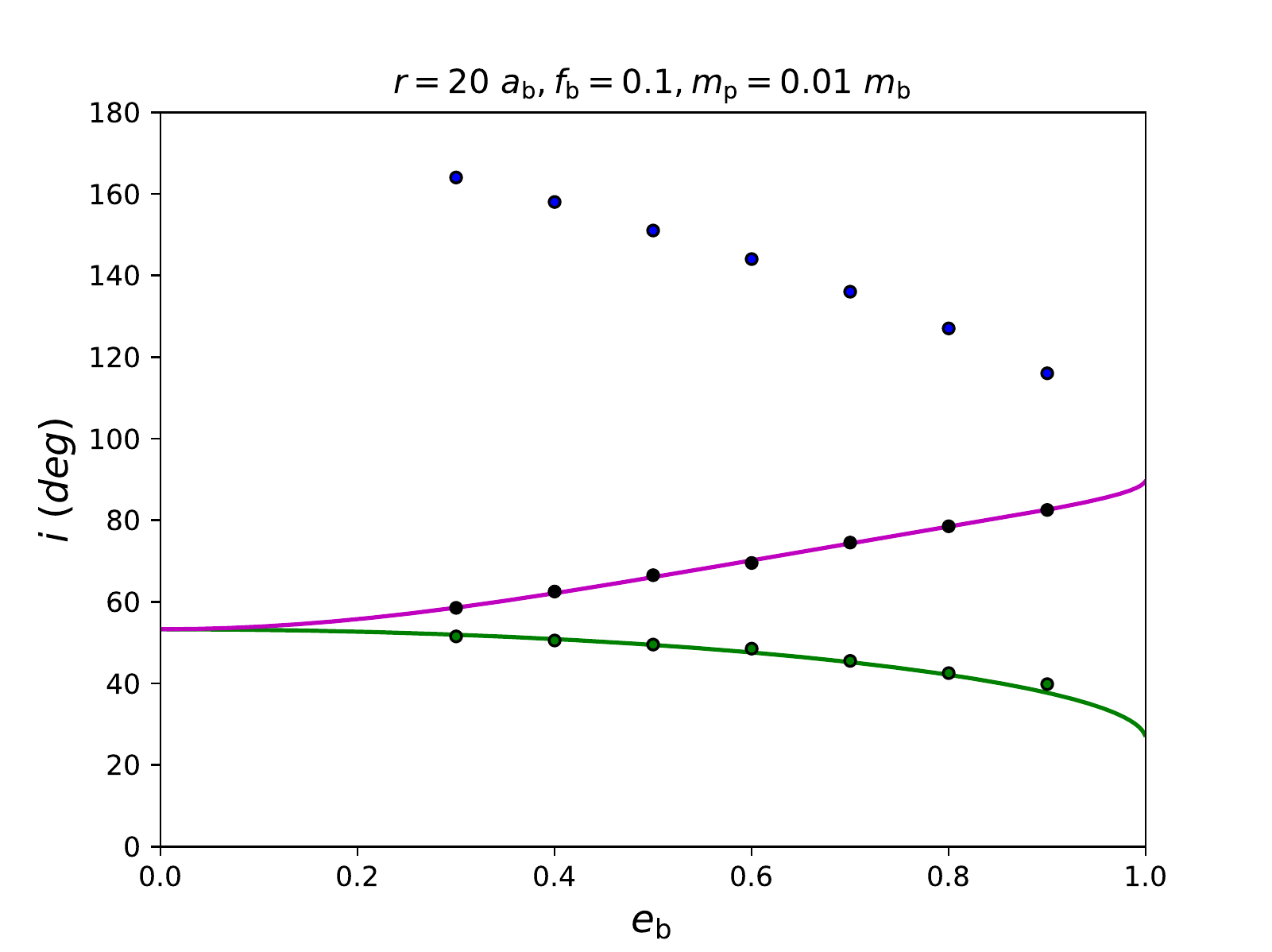}
     \caption{ Same as Fig.~\ref{fig:critical1} except that the planet is orbiting at $r = 20 a_{\rm b}$ and the blue dotted line in the lower right panel shows the boundary between the polar libration and the crescent orbits region for the high mass planet models.  } 
     \label{fig:critical2}
\end{figure*}

\begin{figure}
    \centering
    \includegraphics[width=8.8cm]{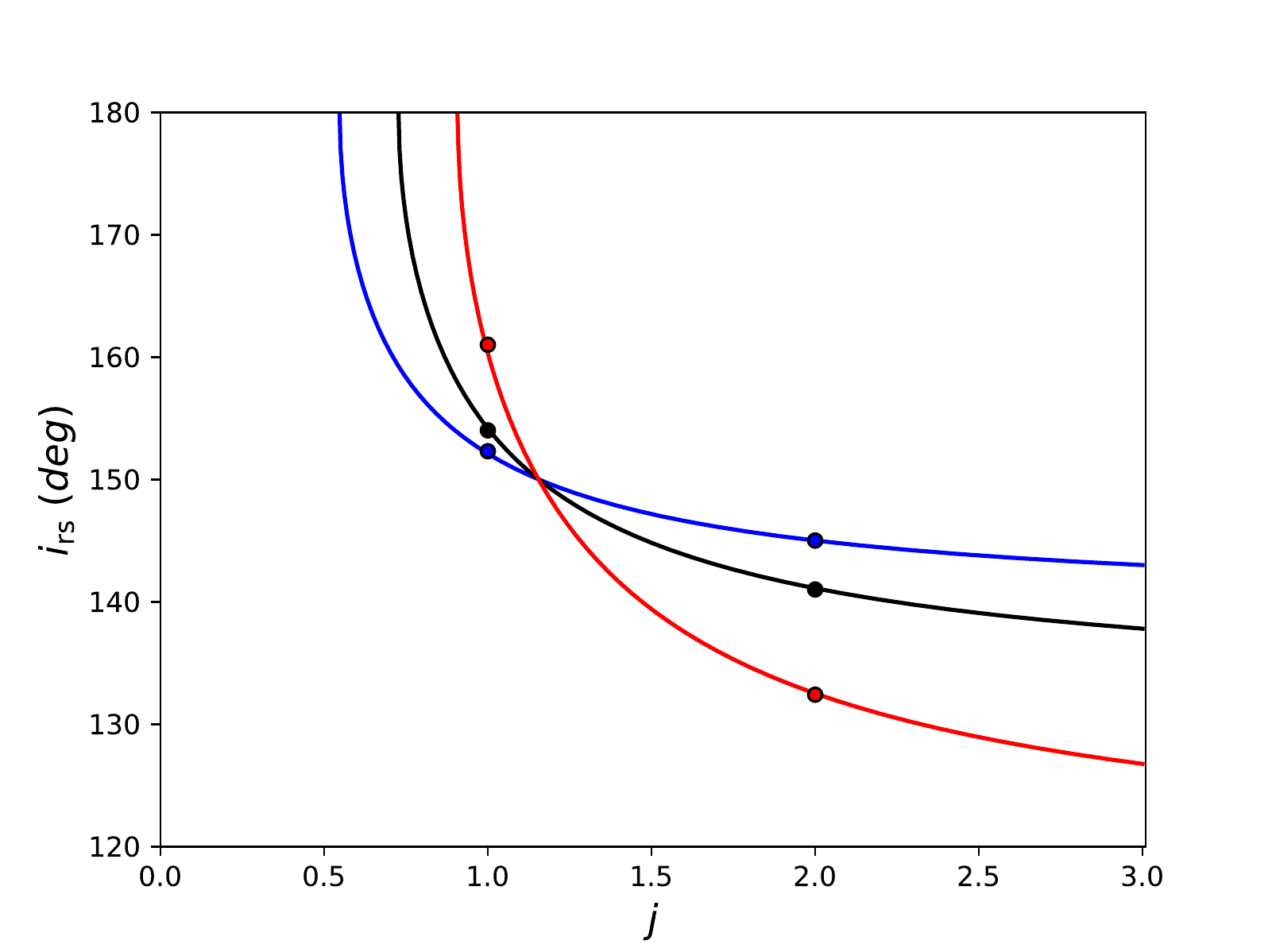}
    \caption{Comparison of the retrograde analytic solution given by Equation (\ref{eq:ic}) with simulation results of Models G1 to H3 for the retrograde stationary tilt $i_{\rm s}$ of the planet relative to the binary as a function of planet-to-binary angular momentum ratio $j$ with binary eccentricity $e_{\rm b}$ = 0.2 (blue line), 0.5 (black line), and 0.8 (red line). The six dots represent the simulation results with $j = 1$ and $j=2$. The curves reach $i_{\rm s}=180^\circ$ at $j=j_{\rm cr}$
    given by Equation (\ref{jcrit}).}
    \label{fig:irc}
\end{figure}

The green and red solid lines in Fig.~\ref{fig:critical1} and Fig.~\ref{fig:critical2} plot the analytic solutions of $i_{\rm min}$ obtained in Equation (\ref{eq:icrit1}) (green segments for $\chi <0$) and Equation (\ref{eq:icrit2}) (red segments for $\chi >0$) as a function of $e_{\rm b}$. We also determined critical angles numerically in our simulations.
The green dotted lines and blue dotted lines in Fig.~\ref{fig:critical1} and Fig.~\ref{fig:critical2} show the simulation results for minimum and maximum inclinations, respectively. Fig.~\ref{fig:critical1} and Fig.~\ref{fig:critical2} are determined for librating orbits of a low mass planet ($m_{\rm p} = 0.001\,m_{\rm b}$) and a high mass planet ($m_{\rm p} = 0.01\,m_{\rm b}$), respectively.  Note that the dotted lines are shorter in the low binary mass fraction plots because some of the orbits are unstable in our simulations. However, the analytic solutions cover the entire range of binary eccentricities. The analytic solutions and the three--body simulations are in very good agreement. The comparison shows that both branches of the $i_{\rm min}$ analytic solutions (red and green) agree well with the simulations.

For a low mass planet, the value for $i_{\rm min}$ is rather insensitive to the location of the planet or $f_{\rm b}$ and so the curves look similar in each panel.  This insensitivity can be understood by the fact
that such models have small values of angular momentum ratio $j$. In the limit that $j$ goes to zero, we have from Equation (\ref{chi}) that $\chi >0$ for $e_{\rm b} >0$. Therefore, $i_{\rm min}$ is given by Equation (\ref{eq:icrit1})
and the plotted curves should be nearly entirely red, rather than green. In this limit that $j$ goes to zero, 
Equation (\ref{eq:icrit1}) reduces to
\begin{equation}
    \cos i_{\rm min}= \frac{\sqrt{5} e_{\rm b}}{\sqrt{1+4e_{\rm b}^2}}.
\end{equation}
This equation implies that $i_{\rm min}$ decreases monotonically from $90^\circ$ to $0^\circ$ as $e_{\rm b}$ increases from 0 to 1, as we find in the low mass planet plots. 

%However, for the high mass planet, $i_{\rm min}$ is larger if $e_{\rm b}$ > 0.3 and  smaller if $e_{\rm b}$ < 0.3.  . However, the analytic solutions show the whole range of binary eccentricities. The analytic solution and the three--body simulations are in  very good agreement
In  the limit that $j$ is large, we have that $\chi <0$ in Equation (\ref{chi}) and the entire curve for $i_{\rm min}$
should be green, as we find in the lower right panel of Fig.~\ref{fig:critical2}.
In that limit, Equation (\ref{eq:icrit2}) applies and 
\begin{equation}
\cos i_{\rm min} = \sqrt{\frac{3}{5}},
\label{iminlargej}
\end{equation}
independent of $e_{\rm b}$,
as given by equation~(40) in \cite{MartinandLubow2019}.
This minimum angle for libration is the same as the so-called Kozai-Lidov angle of $\simeq 39.2^\circ$ \citep{Kozai1962,Lidov1962}.

Therefore, in this high $j$ limit, $i_{\rm min}$ should be constant and lie along the $\chi <0$ (green) branch, 
independent of $e_{\rm b}$ for $e_{\rm b} < 1$. In the lower
right panel of Fig.~\ref{fig:critical2} (the highest $j$ panel), 
we find that the green line is roughly what we predict.
(There is a small red region close to $e_{\rm b}=1$ that is not visible in the plot.)
The upper left panel in this figure has a smaller $j$ value for a given $e_{\rm b}$ than in
the lower right panel. In going from the former panel to the latter panel, we see that the behaviour of $i_{\rm min}$
is approaching the expectations of Equation (\ref{iminlargej}).

The lower right panel of Fig.~\ref{fig:critical2} shows the results for the simulations with the same parameters as those in the upper right panel, but with a smaller binary mass fraction.  In the smaller binary mass fraction case, there are no retrograde circulating orbits, only the crescent shaped orbits described in Fig.~\ref{fig:surface3}. Thus, the maximum libration angle (the blue dotted line) shows a different trend compared to the other parameters which have retrograde precessing orbits. 

\section{Discussion and Conclusions}
\label{conclusion}

In this paper, we investigated the orbital evolution of a misaligned circular orbit planet  with nonzero mass around an eccentric orbit binary by means of numerical simulations. The planet and binary interact gravitationally and the orbits of both vary in time. In particular, both undergo nodal precession in the inertial frame.
In our suite of three-body simulations, we consider a low mass planet with $m_{\rm p} = 0.001 m_{\rm b}$ at $r=5\,a_{\rm b}$ and a high mass planet with $m_{\rm p} = 0.01 m_{\rm b}$ at $r=5\,a_{\rm b}$ and at $r=20\,a_{\rm b}$ along with some even higher angular momentum third bodies.
We considered different values of the eccentricity of the binary, $e_{\rm b}$, its mass fraction, $f_{\rm b}$, and the planet's initial inclination  $i$. To map out the possible orbits in these systems, we concentrated on numerically determining
the transitions between the orbit families (circulating and librating for both prograde and retrograde orbits). In addition, we determined the stationary
orbits for which the relative tilt and nodal phase between the planet orbit and binary orbit  are constant in time.

For a very small planet mass, there are two stationary orbital states:
coplanar and polar. In the polar state, the stationary planet--to-binary tilt is $90^\circ$ and the angular momentum of the planet
is along the binary eccentricity vector. The stationary states in the case that the planet mass is nonzero is a generalisation to the polar state, but with the relative orientation not being perpendicular and the nodal phase not being constant in time in the inertial frame.

 Equations (\ref{eq:ic}), (\ref{chi}), (\ref{eq:icrit1}), 
and (\ref{eq:icrit2}) predict that that the only parameters that control 
the planet--to--binary tilt for the stationary orbit and the minimum tilt for the transition from circulation to libration are 
the binary eccentricity $e_{\rm b}$ and the ratio of the planet--to--binary angular momentum $j$. 
Our numerical results agree with this prediction.
Other parameters, such as the binary mass fraction $f_{\rm b}$, only cause
changes in these angles through their dependence on $e_{\rm b}$ or $j$. For example, in Fig.~\ref{fig:ac} we see that the stationary angle depends only on $j$ for fixed $e_{\rm b}$ for different values of binary mass fraction $f_{\rm b}$. 
In addition, the general agreement between the simulations and analytic predictions
across a range of parameters implies that this dependence holds. These angles are also related to the evolution of discs. Simulations by \cite{MartinandLubow2019} suggest that a prograde disc approaches the stationary angle given by Equation (\ref{eq:ic}), if we consider $j$ to represent the disc--to--binary angular momentum ratio.

The numerical results agree very well with the analytic equations for the stationary and minimum libration tilts given in \cite{MartinandLubow2019}
(see Figs.~\ref{fig:ac}, \ref{fig:critical1}, \ref{fig:critical2}, and \ref{fig:irc}). These analytic equations are based on the quadrupole
approximation for the secular binary gravitational field \citep{Farago2010}.
We find numerically that this approximation holds well, even for orbits
that are fairly close to the binary $\sim 3 a_{\rm b}$ that is near the
orbital radius where instability sets in (see Fig.~\ref{fig:inner}).

As predicted analytically, the main effect of increasing angular momentum ratio $j$ for fixed $e_{\rm b}$ is to monotonically decrease the relative planet--to--binary stationary tilt $i_{\rm s}$ in the prograde case (where $i_{\rm s} \le 90^{\circ}$) (see Fig.~\ref{fig:ac}). In addition, this stationary tilt increases with
increasing $e_{\rm b}$ for fixed $j$ in the prograde case.

The behaviour of the stationary tilt in the retrograde case is more complicated, but agrees with the analytic predictions given in  \cite{MartinandLubow2019}.
In this case, the stationary inclination for noncoplanar orbits decreases
with increasing $j$ for fixed $e_{\rm b}$ as in the prograde case. But $i_{\rm s}$
changes from increasing with $e_{\rm b}$ to decreasing at $j=2/\sqrt{3}$.
In addition, there are no noncoplanar stationary orbits below a certain
$j$ value, denoted by $j_{\rm cr}$, that depends on $e_{\rm b}$ (see Fig.~\ref{fig:irc} and Equation
(\ref{jcrit})). This property does not hold in the prograde case. 

Another difference between the prograde and retrograde cases is the
topology of librating orbits. In the prograde case, the librating orbits
are always nested in a $i\cos \phi -i\sin \phi$ phase plane about point with $i=i_{\rm s}$ and $\phi=\pm 90^\circ$ (e.g., red and cyan lines in
Fig.~\ref{fig:surface1}). In the retrograde case, for $j> j_{\rm cr}$
described above, librating orbits are not fully nested within each other
and do not orbit about
the stationary point in the $i\cos \phi -i\sin \phi$ phase plane
(e.g., magenta lines in Fig.~\ref{fig:surface4}). In this phase plane, prograde librating orbits have an oval shape,
while retrograde librating orbits have a crescent shape.

The variation of $e_{\rm b}$ in time is significantly larger for a higher mass planet. In the simulation with  initial conditions $e_{\rm b} = 0.2$, $f_{\rm b}= 0.1$ and $i = 170^{\circ}$, the binary eccentricity can be excited to values very close to 1 (see Fig.~\ref{fig:surface1}). This behaviour is similar to what occurs with Kozai-Lidov oscillations in which the outer object is a planet \citep[e.g.][]{Naoz2016}.

The recent release of data from {\it Gaia} allowed us to better characterize the kinematics of the eccentric equal mass close binary HD 106906 in the Lower Centaurus Crux group \citep{Bailey2014}. This system hosts both a wide asymmetric debris disc and a planetary-mass companion. The formation mechanism of the planet and the stability of its orbit are still under debate \citep{Rodet2017, Derosa2019}. The binary components have masses $M_1=1.37\,\rm M_\odot$ and $M_2=1.34\,\rm M_\odot$. The orbital period is $49.2\,\rm day$ and the binary eccentricity is $e=0.67$ \citep{Derosa2019}.  The planet, HD 106906b was directly imaged with a projected separation of 738 au and was found to be oriented at $21^{\circ}$ from the position angle of the disc midplane, suggesting that the planet's orbit is not coplanar with the system \citep{Kalas2015}.  The planet mass was inferred to be $m_{\rm p}=11\pm 2\, M_{\rm J}$ \citep{Bailey2014}. While the orbital properties of the planet are uncertain, if we assume that the semi-major axis is $738\,\rm au$ and the eccentricity is zero, the ratio of the angular momentum of the planet to the angular momentum of the binary is $j=1.13$. According to our analytical calculations, the stationary inclination is $i_{\rm s}=73.5^{\circ}$ and the  minimum critical angle between the prograde circulating and the librating orbit regions is $i_{\rm min}=41.6^{\circ}$.  

The expected orbital properties of a high inclination circumbinary planet depend
on the mass of the gas disc in which it forms, and when the planet forms. \cite{MartinandLubow2019} showed that prograde discs
of high mass align to the generalised polar state at tilts that are less than $90^\circ$. A giant planet that formed in such a disc would be expected to open a gap and not remain coplanar with the disc \citep{Lubow2016,Pierens2018}. Such a massive planet would undergo libration oscillations in its orbit about an
inclination of less than $90^\circ$, even after the disc has dispersed. As the disc loses mass, the inclination of the generalised polar state moves closer towards $90 ^\circ$. Thus, the timescale of the disc dispersal may affect the final inclination of debris left over from the gas disc. A low mass gas disc, or massive disc that is dispersed on a sufficiently long timescale forms  a debris disc that orbits close to  polar, as is the case for 99 Herculis, that is $3^\circ$ away from polar alignment \citep{Kennedy2012}. A planet (e.g., an Earth-like planet) formed from the resulting polar debris disc would lie on a stationary polar orbit.  A debris disc that is not close to polar or coplanar would be subject to violent collisions due
to nodal differential precession. 

\section*{Acknowledgements}
 Computer support was provided by UNLV's National Supercomputing Center. C.C. acknowledges support from a UNLV graduate assistantship. We acknowledge support from NASA through grants NNX17AB96G and 80NSSC19K0443.

%\bsp	
\label{lastpage}
\end{document}